\documentclass[a4paper,fleqn,usenatbib]{mnras}

\usepackage{newtxtext,newtxmath}

\usepackage[T1]{fontenc}
\usepackage{ae,aecompl}

\usepackage{graphicx}	
\usepackage{amsmath}	

\title[Deep learning application of a FUV all sky map]{Construction of a far ultraviolet all sky map from an incomplete survey: Application of a deep learning algorithm}

\author[Y.-S. Jo et al.]{
	Young-Soo Jo$^{1}$\thanks{E-mail: stspeak@kasi.re.kr},
	Yeon-Ju Choi$^{2}$,
	Min-Gi Kim$^{3}$,
	Chang-Ho Woo$^{3}$,
	Kyoung-Wook Min$^{3}$, \newauthor
	and Kwang-Il Seon$^{1,4}$
	\\
	$^{1}$Korea Astronomy and Space Science Institute, 776 Daedeok-daero, Yuseong-gu, Daejeon 34055, Republic of Korea\\
	$^{2}$Korea Aerospace Research Institute, 169-84, Gwahak-ro,Yuseong-Gu, Daejeon, 34133, Republic of Korea\\
	$^{3}$Department of Physics, Korea Advanced Institute of Science and Technology, 291 Daehak-ro, Yuseong-gu, Daejeon 34141, Republic of Korea\\
	$^{4}$Astronomy and Space Science Major, University of Science and Technology (UST), Korea, 217 Gajeong-ro, Yuseong-gu, Daejeon, 34113, Republic of Korea
}

\date{Accepted XXX. Received YYY; in original form ZZZ}

\pubyear{2020}

\begin{document}
	\label{firstpage}
	\pagerange{\pageref{firstpage}--\pageref{lastpage}}
	\maketitle
	
	\begin{abstract}
		We constructed a far ultraviolet (FUV) all sky map based on observations from the Far Ultraviolet Imaging Spectrograph (FIMS) aboard the Korean microsatellite STSAT-1. For the $\sim$20$\%$ of the sky not covered by FIMS observations, predictions from a deep artificial neural network were used. Seven datasets were chosen for input parameters, including five all sky maps of H$\alpha$, E(B-V), N(HI), and two X-ray bands, with Galactic longitudes and latitudes. 70$\%$ of the pixels of the observed FIMS dataset were randomly selected for training as target parameters and the remaining 30$\%$ were used for validation. A simple four-layer neural network architecture, which consisted of three convolution layers and a dense layer at the end, was adopted, with an individual activation function for each convolution layer; each convolution layer was followed by a dropout layer. The predicted FUV intensities exhibited good agreement with Galaxy Evolution Explorer (GALEX) observations made in a similar FUV wavelength band for high Galactic latitudes. As a sample application of the constructed map, a dust scattering simulation was conducted with model optical parameters and a Galactic dust model for a region that included observed and predicted pixels. Overall, FUV intensities in the observed and predicted regions were reproduced well.
	\end{abstract}
	
	
	
	
	\section{Introduction}
	
	Observations of the interstellar medium (ISM) in the far ultraviolet (FUV) wavelengths (900–1750{\AA}) are extremely useful as they provide important information on the physical and chemical processes that occur among the constituents of our galaxy. For example, emission lines of highly ionized atoms provide clues on how Galactic hot gas is generated and cooled due to interaction with the ambient environment \citep{jo19}. The FUV band also contains many molecular hydrogen fluorescence lines, which are good tracers of star-forming regions in molecular clouds \citep{jo17}. Furthermore, because the diffuse continuum background is believed to be the starlight scattered by interstellar dust, its observation helps us understand scattering properties of the dust grains and geometries of molecular clouds, particularly when using scattering simulations of stellar photons \citep{jo12,lim13,cho13,cho15}. FUV continuum intensity has also exhibited good correlation with H$\alpha$ intensity, indicating that the two emissions may have common origins and similar radiative transfer mechanisms \citep{seo11a,seo11b}.
	
	Despite its importance, there have not been many FUV observations in part due to significant extinction effects at FUV wavelengths. There have been only two notable FUV survey missions since the one in 1972 by the TD1 satellite \citep{bok73}. GALEX, launched on 2003 April 28, observed the sky in two wavelength bands using a dichroic mirror: FUV (1350–1750 {\AA}) and near-ultraviolet (NUV: 1750–2850 {\AA}). Its spatial resolution was 5-7$\arcsec$ in the sky over a 1\degr.25 field \citep{mar03,mor07}. The GALEX survey data have been used extensively, especially in regard to the diffuse UV background \citep{mur10, mur14a, mur14b, mur16, hen12, hen15, jyo15, nar17, aks18, aks19}. Though not used frequently, low-resolution spectroscopic data are also available from observations made with an objective grism. Another FUV mission used a Far-ultraviolet IMaging Spectrograph (FIMS), also known as Spectroscopy of Plasma Evolution from Astrophysical Radiation (SPEAR), onboard the spacecraft Science and Technology SATellite-1 (STSAT-1) launched on 2003 September 27 \citep{ede06a,ede06b}. FIMS was a dual-band imaging spectrograph with a short- (S-band: 900–1150 {\AA} with 1.5 {\AA} resolution, 4\degr$\times$5$\arcmin$ field of view) and long-wavelength band (L-band: 1335–1750 {\AA} with 2.5 {\AA} resolution, 8\degr$\times$5$\arcmin$ field of view).Both bands have an imaging resolution of 5\arcmin. GALEX and the FIMS L-band possessed nearly identical spectral bands. Nevertheless, both GALEX and FIMS did not cover the whole sky; GALEX intentionally avoided the observation of bright regions to protect the detector whereas the electrical power problems with STSAT-1 caused frequent interruption and the early termination of FIMS.
	
	A deep artificial neural network (DNN) is a subset of machine learning (ML), composed of multiple interconnected layers of “artificial neurons” with powerful optimization algorithms \citep{lec15}. DNN, based on automatic learning from massive data during the training stage, can capture complex non-linear relationships in data by constructing hierarchical internal representations. DNN is especially successful in image processing; specifically, the convolutional neural network (CNN) adopts a feed-forward neural network inspired by studies of visual recognition mechanisms in mammals \citep{raw17}. CNN generally utilizes a multi-layer architecture consisting of multiple feature-extraction steps such as convolution, pooling, and nonlinear activation layers. While CNN, which utilizes only current inputs, is primarily suitable for image processing, sequential datasets (in which the next step depends on the previous input data) can be modeled by a recurrent neural network (RNN). The RNN algorithm includes a hidden state memory that summarizes the previous input information. RNN has proven to be particularly successful in natural language processing \citep{lip15}.
	
	ML has become increasingly popular in astronomy studies, where enormous multi-spectral datasets are produced from ground and space observations, and very sophisticated analysis tools are required to extract meaningful information \citep{bal10}. For example, CNN was applied to studies of star-galaxy distinction \citep{kim17}, gravitational lensing \citep{sch18,dav19}, and photometric redshifts \citep{pas19}. \citet{nau18} presented an unsupervised auto encoding RNN that could reconstruct noisy, irregularly sampled astronomical data. An ML technique called Support Vector Machine (SVM) used an optimal hyperplane in an N-dimensional space to separate classes \citep{kov15,pas18}. \citet{bar19} developed an outlier detection algorithm based on unsupervised learning that could identify unexpected new objects in image, spectral, and time-series datasets. \citet{das19} estimated the age, distance, and mass of red giant stars using Bayesian artificial neural networks.
	
	Both GALEX and FIMS did not cover the whole sky; the whole region around the Galactic plane and bright stars were not observed by GALEX, while several hollow areas corresponding to the orbits with a power problem exist in the FIMS dataset. Our objective in this study is to fill the unsurveyed regions of the FIMS map for the total FUV intensity of the FIMS L-band, using all sky surveys completed at other wavelengths and a deep learning algorithm. We use the FIMS dataset for this supervised learning because the high-intensity Galactic plane is missing in the GALEX dataset, while FIMS covers both low and high-intensity regions. As it is difficult to confirm the validity of the predicted FUV intensities of the unsurveyed regions due to lack of observations in the corresponding wavelengths, we do not intend to furnish a precise map, but rather present a reasonable map of a global scale constructed with the assistance of recently-developed neural network techniques. Nevertheless, indirect validation is possible by comparing the map with simulations of dust scattering in FUV which have often been used to model the interstellar dust, as well as with the H$\alpha$ map which is similar to the diffuse FUV map after extinction correction in both wavelengths. Another objective of our study is to demonstrate the utility of machine learning in astrophysics: The present work shows that a very simple deep learning algorithm produces remarkable results from unbiased databases even without imposing any a priori assumptions based on experience. The present paper is organized as follows: Section \ref{sec:const} delineates the construction of the FUV all sky map from the incomplete FIMS data. We first outline the structure of the FIMS data, followed by the detailed description of the DNN algorithm. The resulting map is presented and compared with GALEX observations. In Section \ref{sec:discus}, we apply dust scattering simulations to verify the validity of the predicted FUV intensities. We conclude the paper with a summary in Section \ref{sec:summa}.

	\section{Construction of FUV the all sky map}
	\label{sec:const}
	
	\subsection{Datasets}
	\label{sec:datas} 
	
\begin{figure*}
		\centering
		\includegraphics[clip,scale=0.33,angle=90]{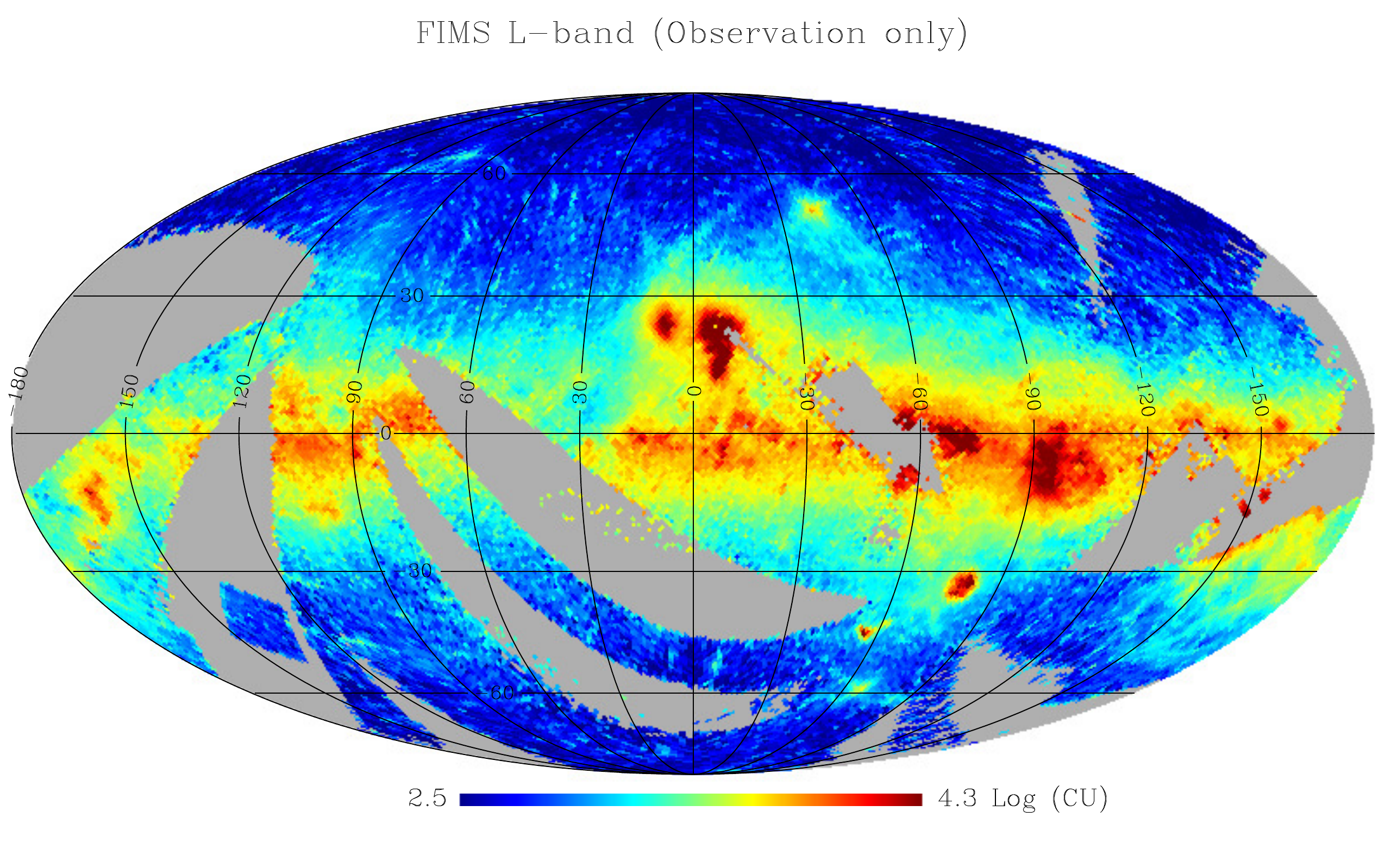} \,
		\includegraphics[clip,scale=0.33,angle=90]{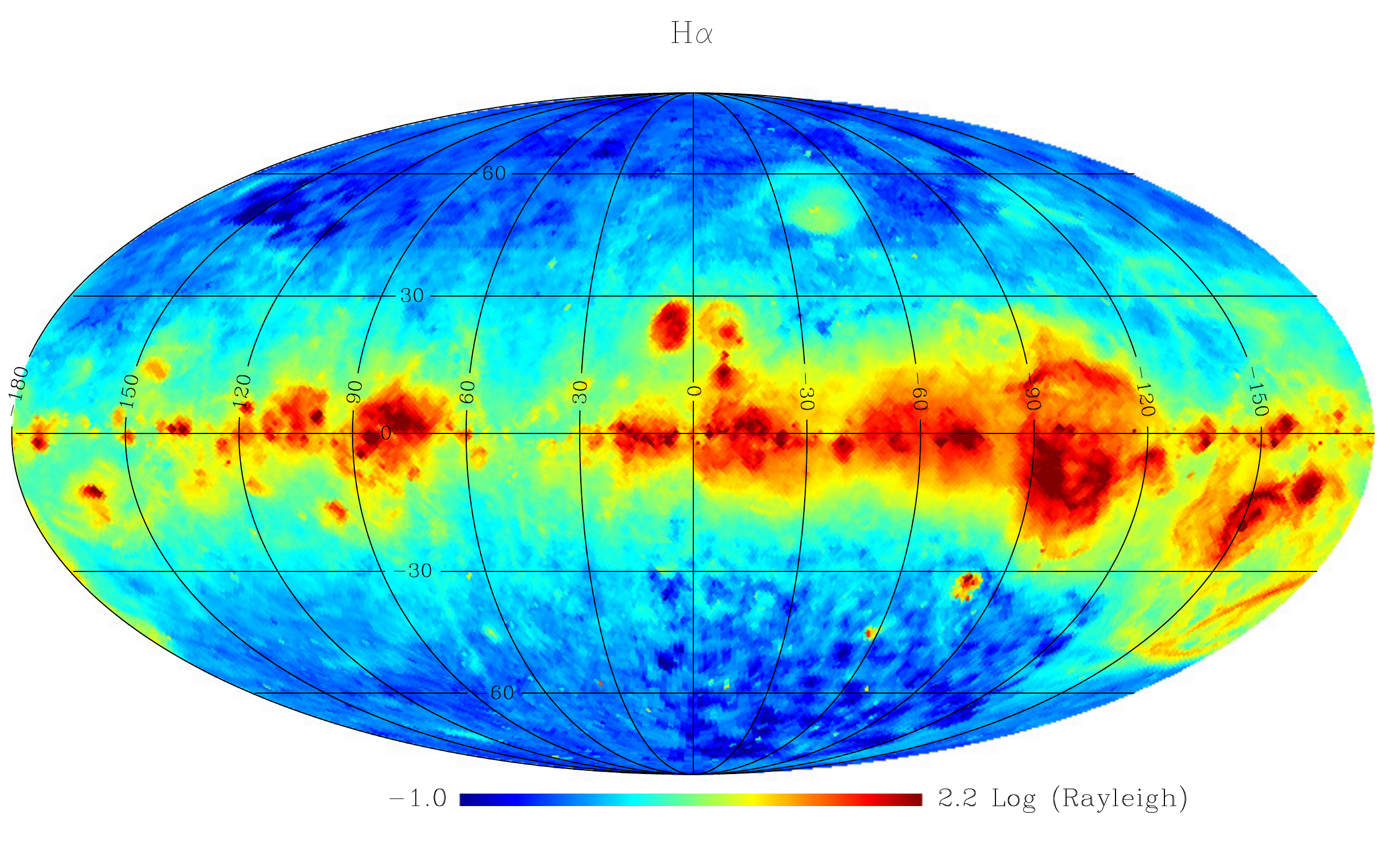} \\
		\includegraphics[clip,scale=0.33,angle=90]{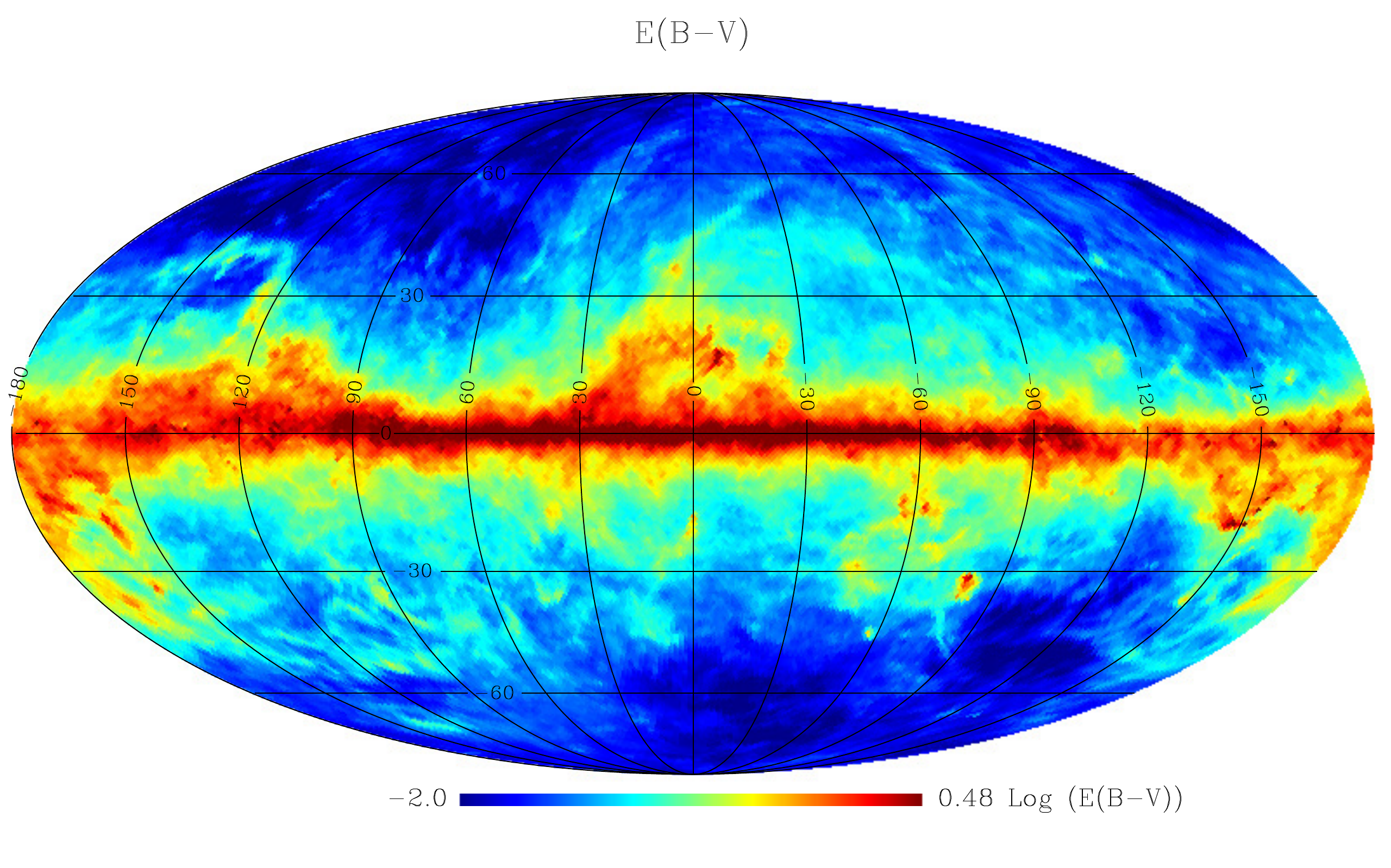} \,
		\includegraphics[clip,scale=0.33,angle=90]{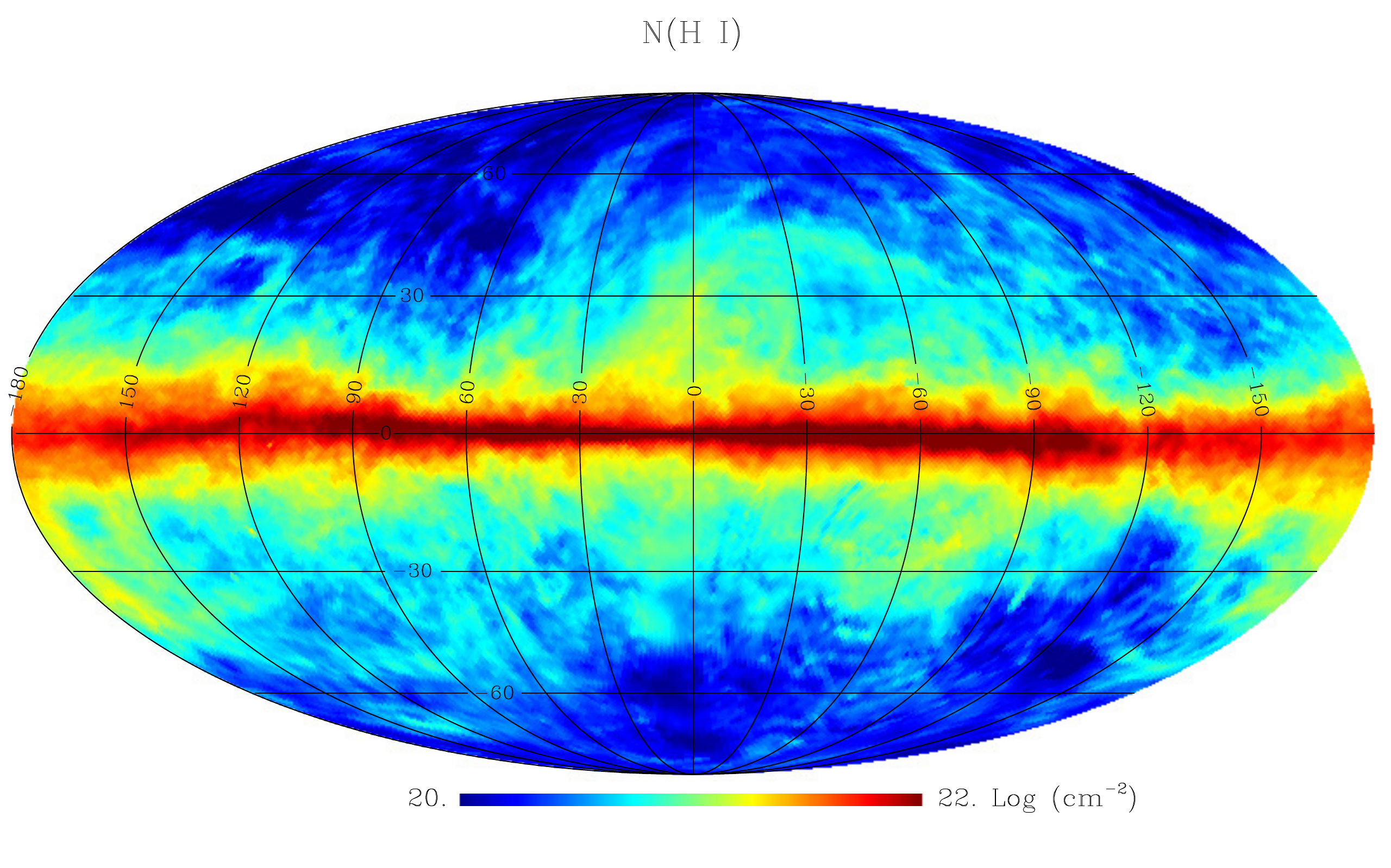} \\
		\includegraphics[clip,scale=0.33,angle=90]{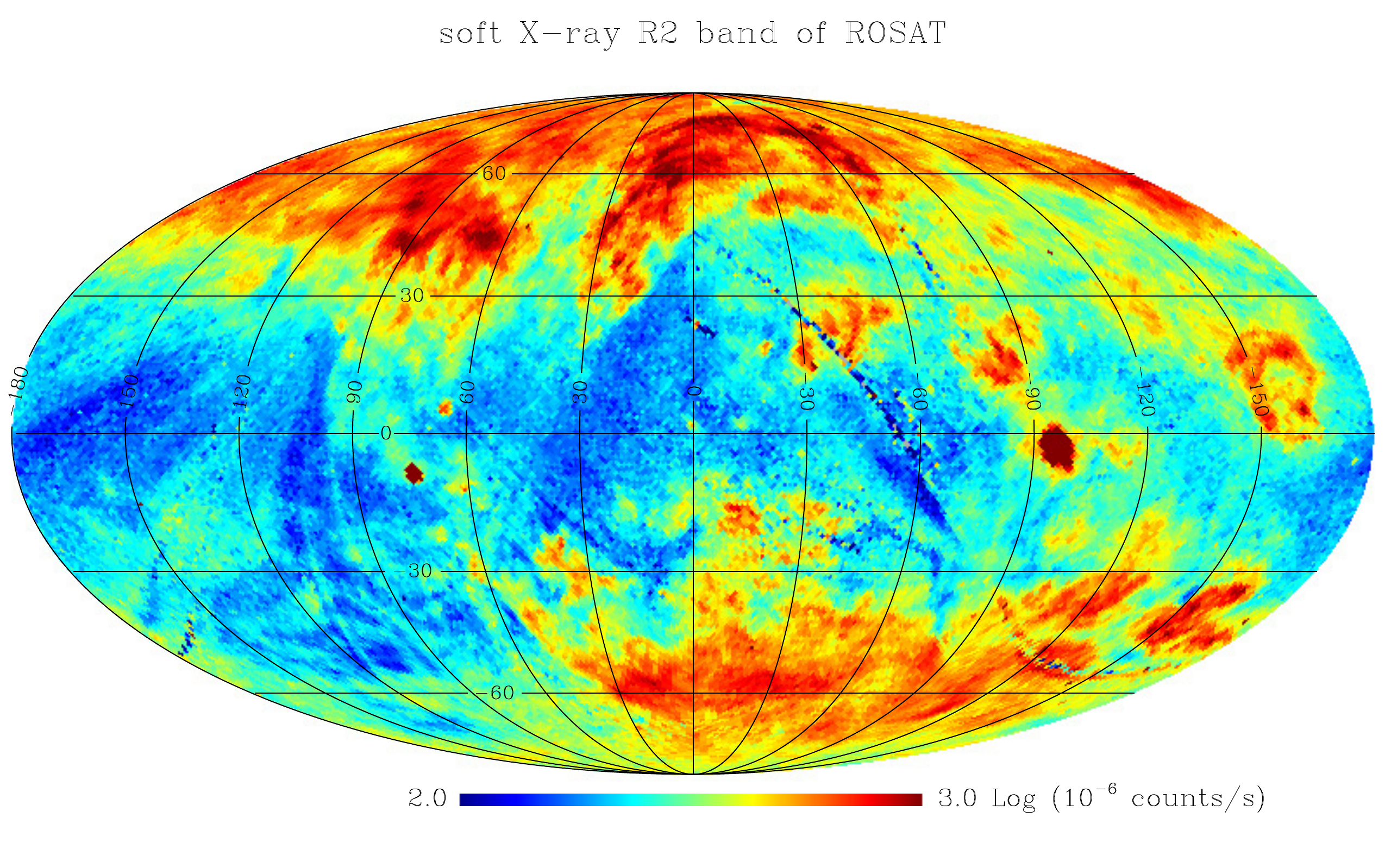} \,
		\includegraphics[clip,scale=0.33,angle=90]{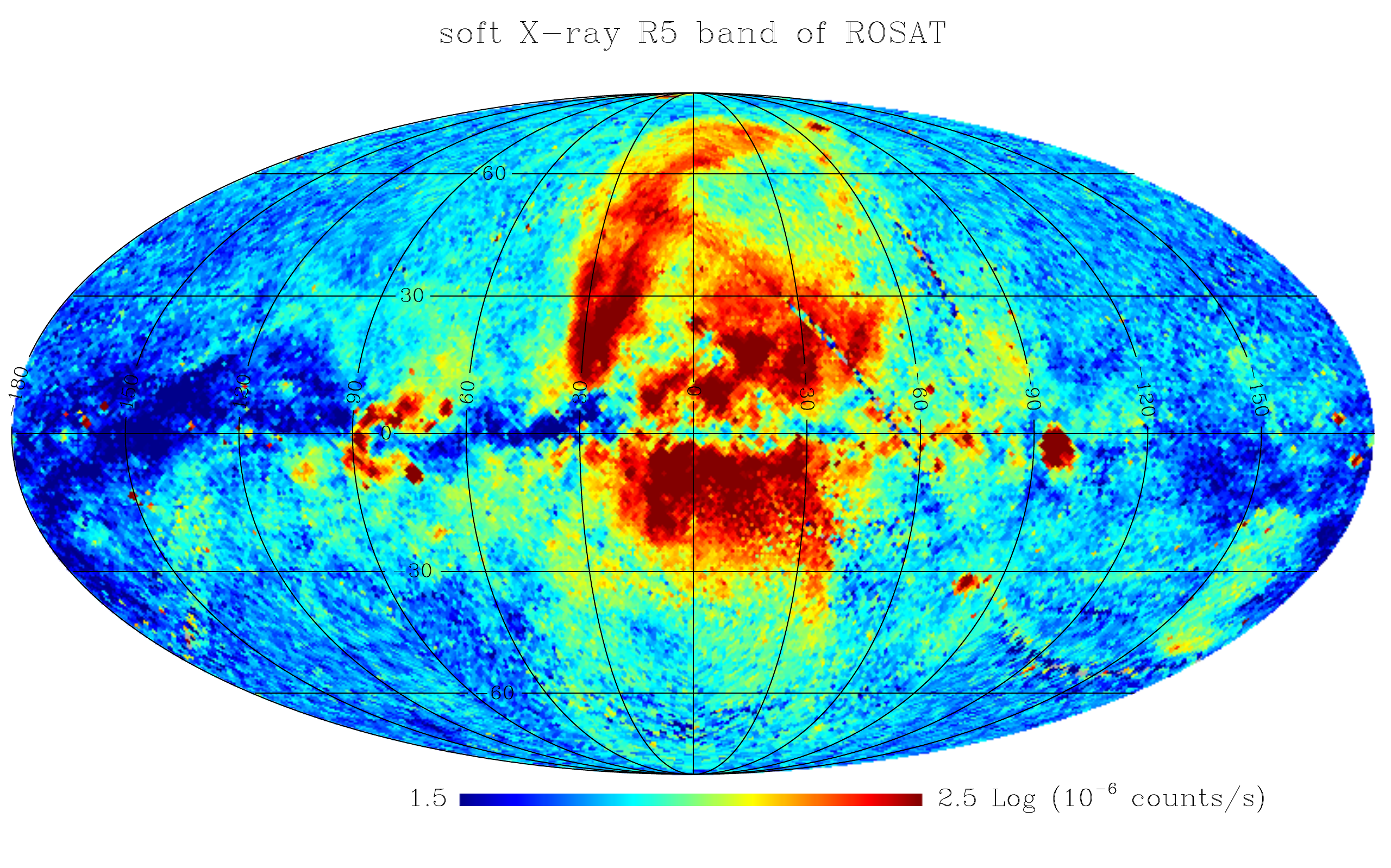} \\
		\caption{All sky maps utilized in the study: (a) FUV observed by FIMS, (b) H$\alpha$, (c) E(B-V), (d) N(HI), (e) soft X-ray R2 band of ROSAT, and (f) soft X-ray R5 band of ROSAT.}
		\label{fig:f01}
	\end{figure*}

	FIMS data that covers only $\sim$75 \% of the sky \citep{seo11a} were stored using the HEALPix scheme \citep{gor05} with a resolution parameter of N$_{\rm side}$=512, corresponding to $\sim$7\arcmin. The diffuse emission map\footnote{\href{https://drive.google.com/file/d/1ZO\_3n7-aUijxLUJA1auhL2l18fRCVars/view?usp=sharing}{https://drive.google.com/file/d/1ZO\_3n7-aUijxLUJA1auhL2l18fRCVars/view?usp=sharing}} of the FIMS L-band was constructed from this archival data with a resolution parameter of N$_{\rm side}$=64 (corresponding to $\sim$1\degr), after the bright pixels corresponding to bright stars had been removed \citep{jo17}. The angular resolution of $\sim$1\degr, which is much lower than the original archival dataset, was selected to secure the minimum signal-to-noise ratio to be above 5 even for the faintest pixel of $\sim$200 CU. With this resolution, the entire sky consisted of 49,152 total spatial pixels, 37,152 of which had non-zero values obtained by FIMS and 12,000 of which were empty. Fig. \ref{fig:f01} shows the FIMS diffuse FUV emission map of the L-band with a resolution of $\sim$1\degr.

	To estimate the FUV intensities for regions not covered by FIMS, the following survey data obtained for other wavelengths were used as input data throughout the training, validation, and prediction stages: A composite H$\alpha$ map\footnote{\href{https://lambda.gsfc.nasa.gov/data/foregrounds/halpha/lambda\_halpha\_fwhm06\_0512.fits}{https://lambda.gsfc.nasa.gov/data/foregrounds/halpha/lambda\_halpha\_fwhm06\_0512.fits}} of \citet{fin03}; E(B-V) map\footnote{\href{https://lambda.gsfc.nasa.gov/data/foregrounds/SFD/lambda\_sfd\_ebv.fits}{https://lambda.gsfc.nasa.gov/data/foregrounds/SFD/lambda\_sfd\_ebv.fits}} of Galactic reddening by \citet{sch98}; neutral hydrogen column density N(HI) map\footnote{\href{https://lambda.gsfc.nasa.gov/data/foregrounds/combined\_nh/lambda\_combined\_nh.fits}{https://lambda.gsfc.nasa.gov/data/foregrounds/combined\_nh/lambda\_combined\_nh.fits}} of \citet{dic90}; R2 (1/4 keV) and R5 (3/4 keV) band maps\footnote{\href{http://www.jb.man.ac.uk/research/cosmos/rosat/RASS\_SXRB\_R2.fits}{http://www.jb.man.ac.uk/research/cosmos/rosat/RASS\_SXRB\_R2.fits}}$^,$\footnote{\href{http://www.jb.man.ac.uk/research/cosmos/rosat/RASS\_SXRB\_R5.fits}{http://www.jb.man.ac.uk/research/cosmos/rosat/RASS\_SXRB\_R5.fits}} of the ROSAT constructed by \citet{sno97}. These survey maps are the representative ones ranging from X-ray to radio wavelengths. They trace different ISM phases according to their physical emission mechanism (X-ray originating from hot gas with T $\sim$ 10$^6$ K, H$\alpha$ from photoionized gas with T $\sim$ 10$^4$ K, E(B-V) tracing dust extinction, and N(HI) tracing the neutral atomic hydrogen gas) and show their own distinct shapes, as depicted in Fig. \ref{fig:f01}. We used two X-ray bands because the all-sky maps constructed from these two bands are quite different from each other; the R2 band shows a striking negative correlation with the neutral ISM while the R5 band has an overall positive correlation. We did not attempt to give weights to particular wavelengths as such a biased action reflects our prejudice regarding the detailed emission and extinction physics that depends on wavelengths and varies from region to region, which we are not certain about. We rather left the deep learning algorithm to find out the nonlinear relationships between the FUV intensities and those of other wavelengths. The acquired nonlinear relationship in the learning stage is verified at the validation stage, in which the predicted results are compared with the observed data which were not used in the learning stage. All maps, although they may have high signal-to-noise ratios at higher resolution, were binned to match the resolution of FIMS because the deep learning algorithm requires the input datasets to have pixel-to-pixel correspondence for direct comparisons. Galactic longitudes and latitudes were also employed as input data to account for the spatial variations in the relationship between the input datasets.
	
	We randomly selected 70\% of the 37,152 pixels with non-zero values for training, while the remaining 30\% were used for validation. As each dataset has its own scales that differ from each other by several orders of magnitude, before applying the DNN algorithm we adjusted the dataset ranges by taking logarithmic values and normalizing them such that all final datasets have a zero mean with a standard deviation of 1. This normalization process is necessary because the difference in intensity among the observed datasets is enormous in the present work, by a factor of up to 10$^{21}$. Unscaled input variables may lead to slow or unstable learning, often causing the learning process to fail \citep{sol97}. Therefore, it is a general practice to normalize the input data to have the same dispersions to guarantee convergence. Furthermore, the log-scale normalization process also accords well with the log-normal property found in the ISM density distribution \citep{ost01, seo12, bur15}. The predicted FUV intensities were finally obtained by applying the inverse of the normalization process .

	\begin{figure*}
		\centering
		\includegraphics[clip,scale=0.3]{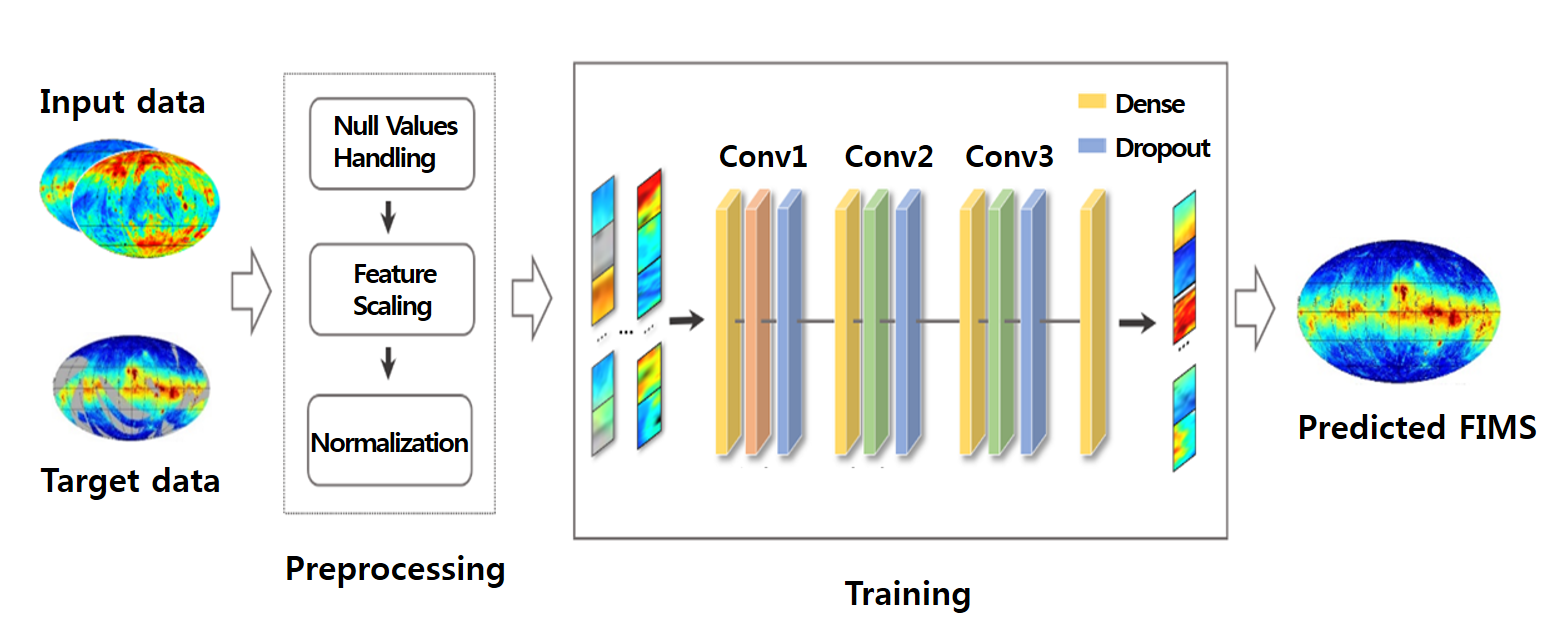}
		\caption{Architecture of the neural network adopted in the present study}
		\label{fig:f02}
	\end{figure*}
	
	\begin{figure*}
		\centering
		\includegraphics[clip,scale=0.5,angle=90]{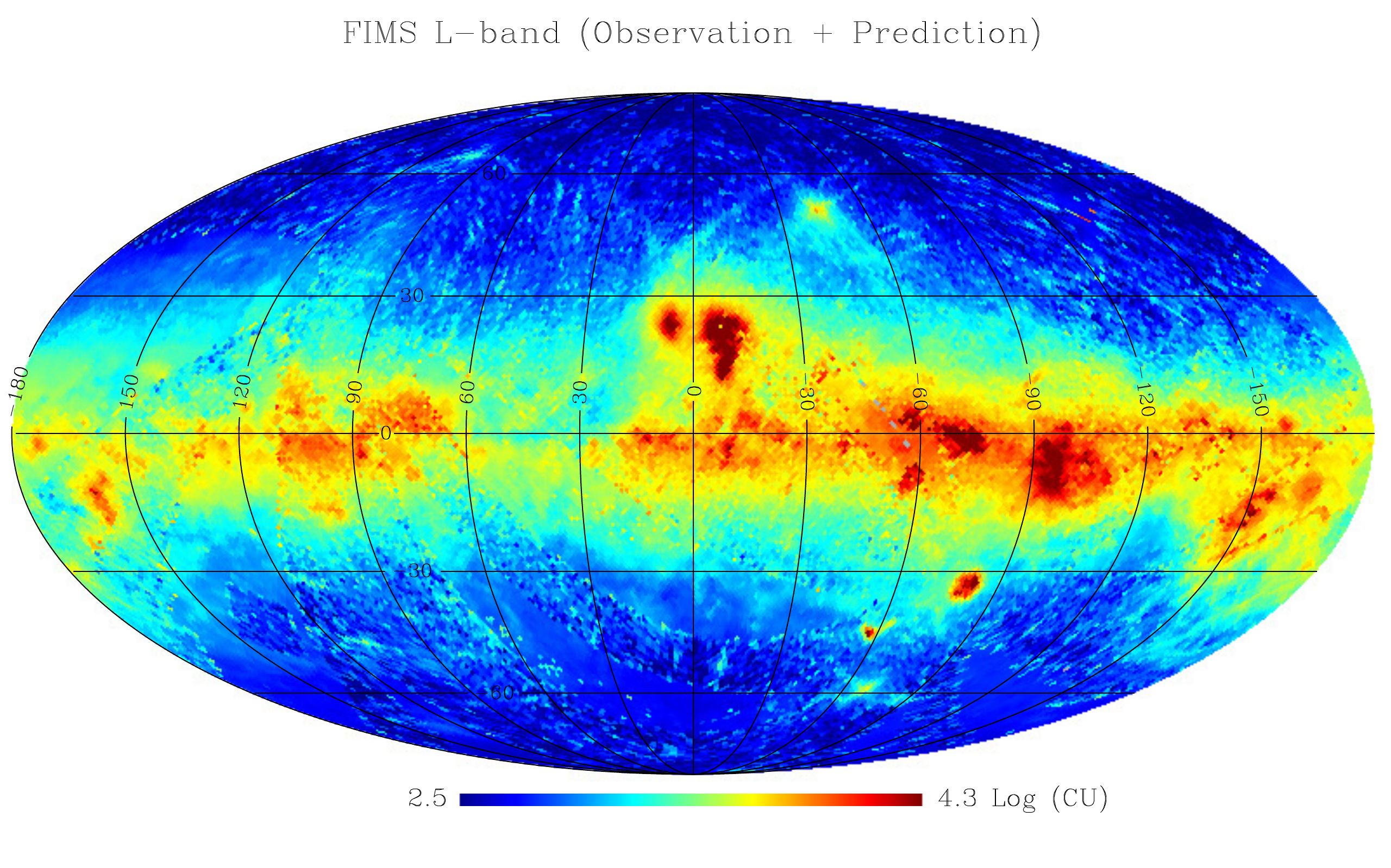}
		\caption{Combined map of the observed and predicted FIMS intensities.}
		\label{fig:f03}
	\end{figure*}

	\subsection{Algorithm}
	\label{sec:algo}
	
	The neural network (NN) generally consists of an input layer for input data, a number of hidden layers for processing and learning, and an output layer. As described in Section \ref{sec:datas}, we split the dataset into training and cross-validation sets before training the model. In the training stage, weights were applied to all input data through a series of layers, and finally the predicted values in the output layer were compared to the target values through a loss function. For the input data, we chose seven datasets, five of which were the observed maps of H$\alpha$, E(B-V), N(HI), and two X-ray bands, and the remaining two corresponds to Galactic longitudes and latitudes, as described in Section \ref{sec:datas}. The FIMS FUV intensities were regarded as the training target parameters. 
	
	We adopted a simple four-layer neural network architecture that consists of three convolution layers and one dense layer at the end, as illustrated in Fig. \ref{fig:f02}. The model has seven input parameters and 16 neurons in every hidden layer; the final dense layer has one predicted output value. For the activation functions, which feature a nonlinear relationship between the input and target datasets, we used a hyperbolic tangent \citep{nwa18} for the first layer and a leaky ReLU \citep{maa13} for the second and the third layers. A hyperbolic tangent is a saturated activation function that generates saturated values for large input parameter values, while the leaky ReLU is non-saturated neurons that provide large output values for large input parameters, and small negative slope instead of being zero when input < 0. We tried several other activation function combinations, but the chosen one provided the best results. We employed a dropout layer after each convolution layer during the training stage to improve errors and reduce overfitting \citep{sri14}. The dropout rates were 0.03, 0.03, and 0.05 for the respective three layers. The present study adopted Keras 2.3.1 library and TensorFlow 1.14.0 as a backend framework \citep{cho15}.

	Internal learnable parameters such as weights and biases were updated using RMSprop \citep{hin12}, an optimization algorithm that utilizes the root mean square of the recent gradients for each weight, with a learning rate of 10$^{-4}$. A random uniform function was adopted for initial weights because completely random initial weights may impair the final predicted values. During the training stage, predicted FUV intensities were compared to observed FIMS values through a loss function that evaluated numerical distances between estimated and actual values. In our model, we used the mean-squared-error (MSE) loss function. Our network was trained with 10$^3$ update iterations and the final loss value was 0.11. Training data and codes can be downloaded via the GitHub link of “https://github.com/yjchoi83/FIMS\_FUV”. FIMS data were compared with GALEX data to confirm the reliability of the final predicted FIMS data. Confirmation tests are described in Section \ref{sec:map}.

	\subsection{FUV all sky map}
	\label{sec:map}
	
	Fig. \ref{fig:f03} shows a combined map of the original observed FIMS data and the predicted FIMS FUV intensities filled in for the unsurveyed regions. It is remarkable that the predicted regions are largely indistinguishable from the observed regions, although the features in the predicted regions (e.g., in the northern region of \textit{l} = 150\degr) are smoother than those in the observed regions. Some bright regions in and around the Galactic plane are evident; for example, the region of (\textit{l}, \textit{b}) $\simeq$ (-50\degr, 0\degr) corresponds to the Lower Centaurus Crux (the nearest OB association), and the region of (-180\degr < \textit{l} < -120\degr, -30\degr < \textit{b} < 0\degr) corresponds to the upper part of the Orion-Eridanus Superbubble that includes Orion nebula, Barnard’s loop, and Lambda Orionis.

	\begin{figure}
		\centering
		\includegraphics[width=\columnwidth]{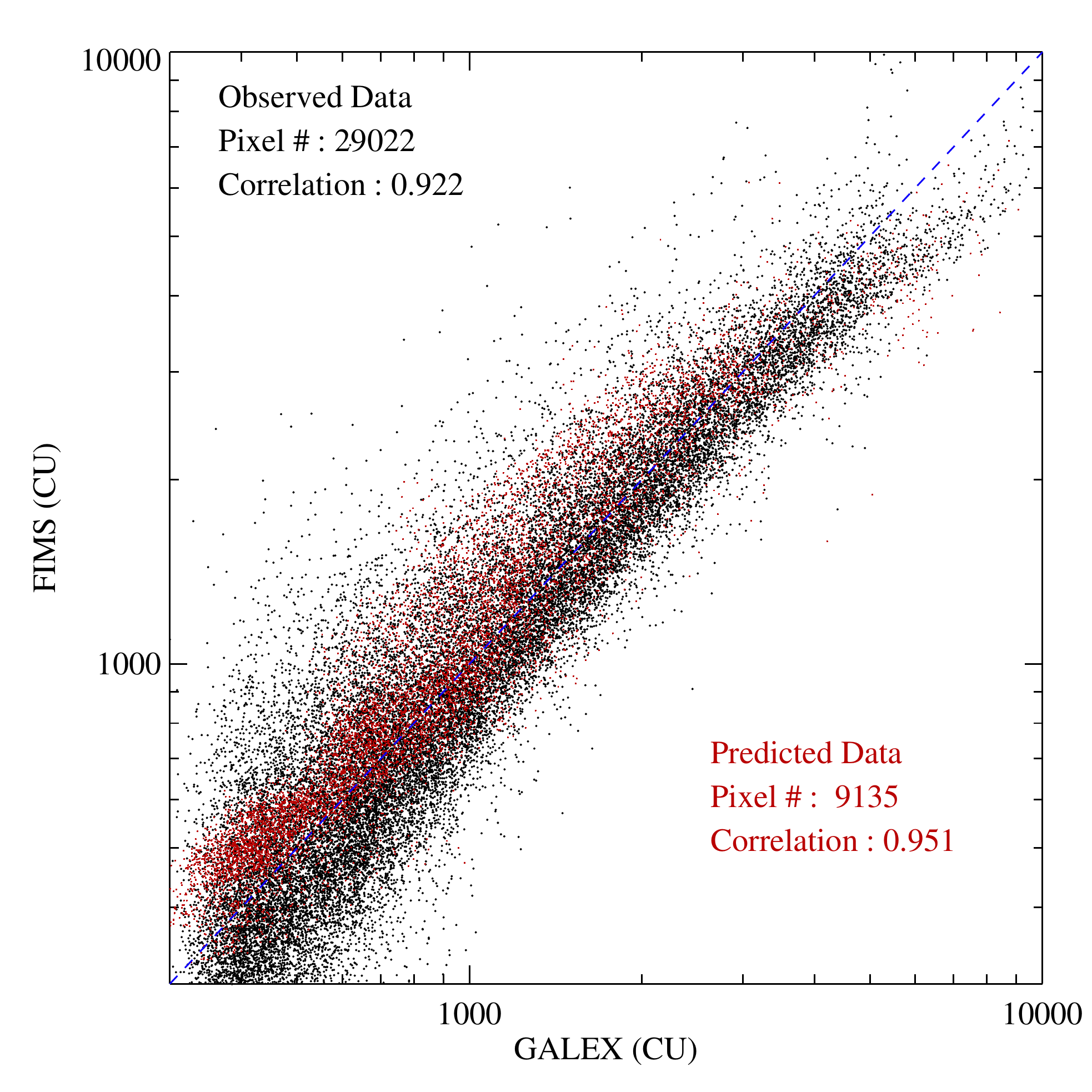}
		\caption{Comparison of the observed and predicted FIMS intensities with GALEX intensities. Black (red) dots represent the observed (predicted) FIMS data. The number of pixels and correlation coefficients estimated using the logarithmic scale are displayed in the upper-left and lower-right corners using the corresponding color notation.}
		\label{fig:f04}
	\end{figure}
	
	\begin{figure}
		\centering
		\includegraphics[width=\columnwidth]{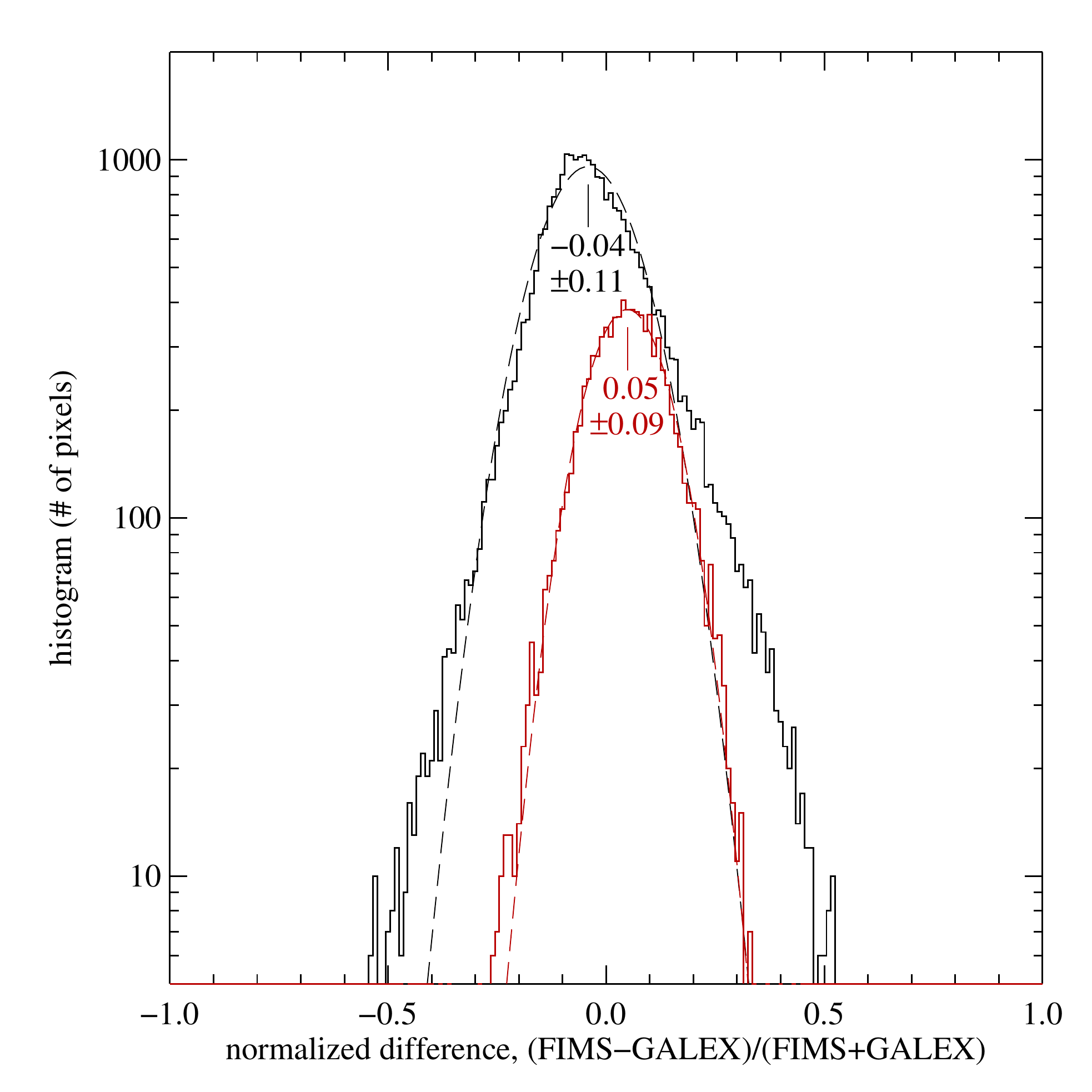}
		\caption{Histogram of the difference between FIMS and GALEX intensities. Black and red histograms correspond to the observed and predicted FIMS data, respectively. Both histograms are fitted using Gaussian curves with centers and widths indicated.}
		\label{fig:f05}
	\end{figure}

	In Fig. \ref{fig:f04}, we compare observed (black) and predicted (red) FIMS intensities with the GALEX observations. Overall, there is a good linear relationship between both the observed and predicted FIMS intensities with the GALEX observations, although the observed FIMS data are somewhat scattered toward higher intensities than the corresponding GALEX data, which resulted in slightly higher predicted intensities. On the other hand, at high intensities the GALEX observations are a little brighter than FIMS. The correlation coefficient estimated using the logarithmic scale exceeds 0.9 for the observed and predicted cases.
	
	Fig. \ref{fig:f05} displays a histogram of the difference between the FIMS and GALEX intensities ((I$_{\rm FIMS}$-I$_{\rm GALEX}$)/(I$_{\rm FIMS}$+I$_{\rm GALEX}$)). Black and red histograms correspond to the observed and predicted FIMS data, respectively. Overall, both histograms are well fitted to single Gaussian functions overlaid using dashed lines. Compared to the fitted Gaussian function, the black histogram shows excess on the right-hand side of Fig. \ref{fig:f05}, implying that the observed FIMS intensities are somewhat brighter than the GALEX intensities. This corresponds to the spread of the observed FIMS data toward higher intensities compared to the GALEX data (Fig. \ref{fig:f04}). These differences are caused by instrumental scattering of bright stellar photons by the slit \citep[see][Figure 1]{jo16}. Unlike the observed FIMS data, the predicted FIMS intensities do not exhibit such excess.

	\begin{figure*}
	\centering
	\includegraphics[clip,scale=0.423]{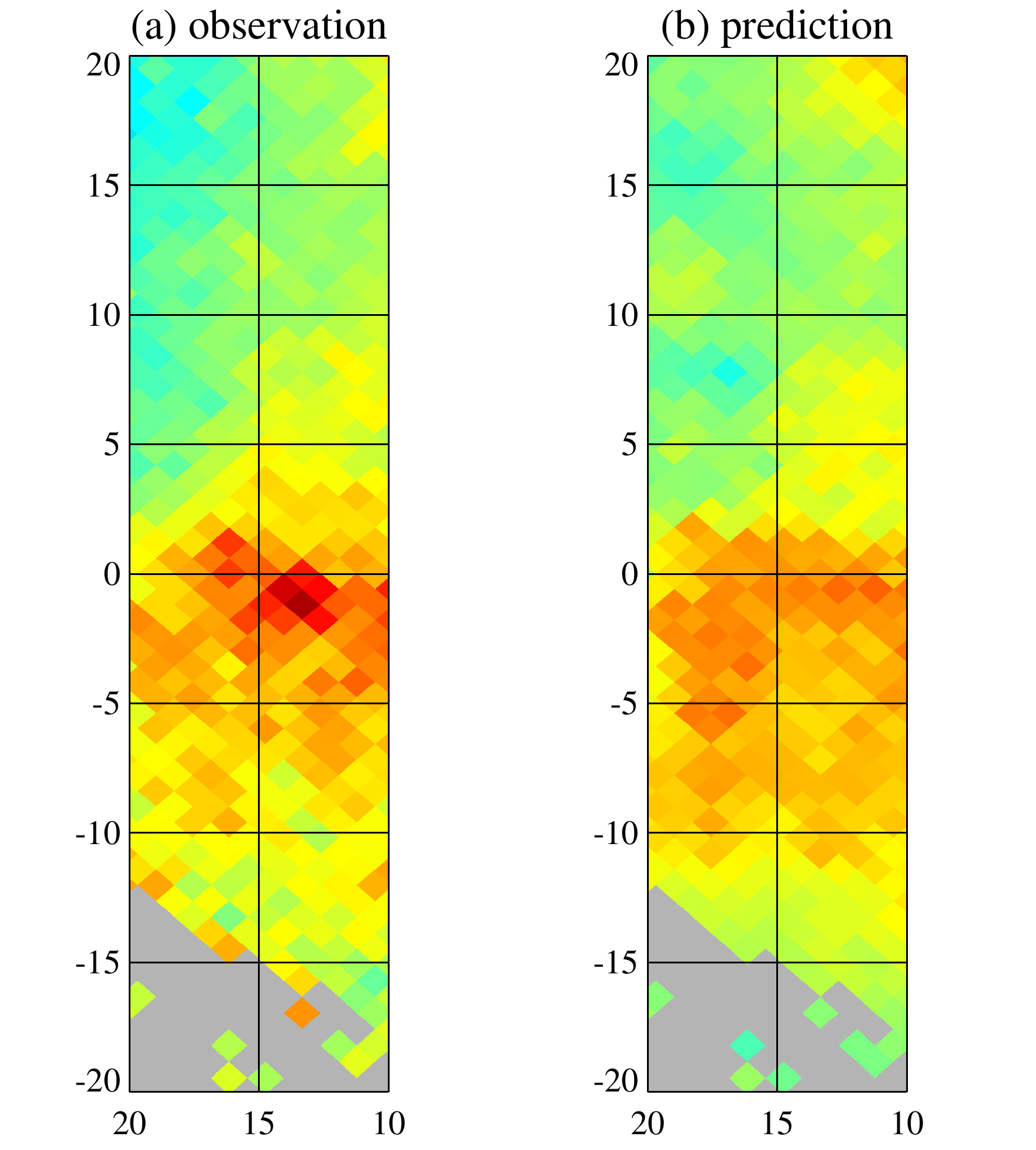} \, \,
	\includegraphics[width=\columnwidth]{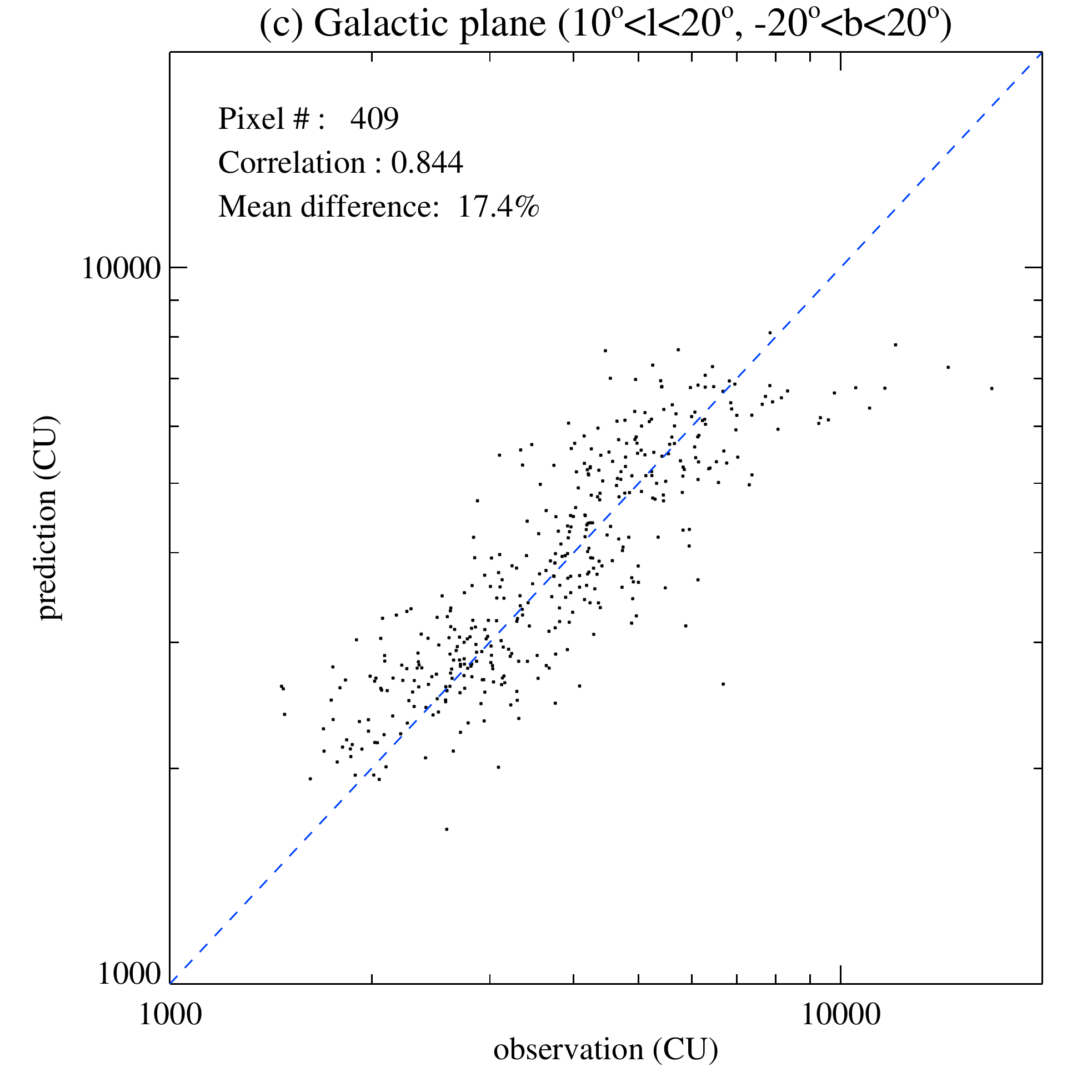}
	\caption{ Comparison of the observed and predicted intensities for the Galactic low latitude region of 10$\degr$ < l < 20$\degr$, |b| < 20$\degr$: (a) actual FIMS observation, (b) predicted result, and (c) scatter plot of the predicted intensities compared with the actual observation. }
	\label{fig:f06}
	\end{figure*}	
	
	\begin{figure*}
		\centering
		\includegraphics[width=\columnwidth]{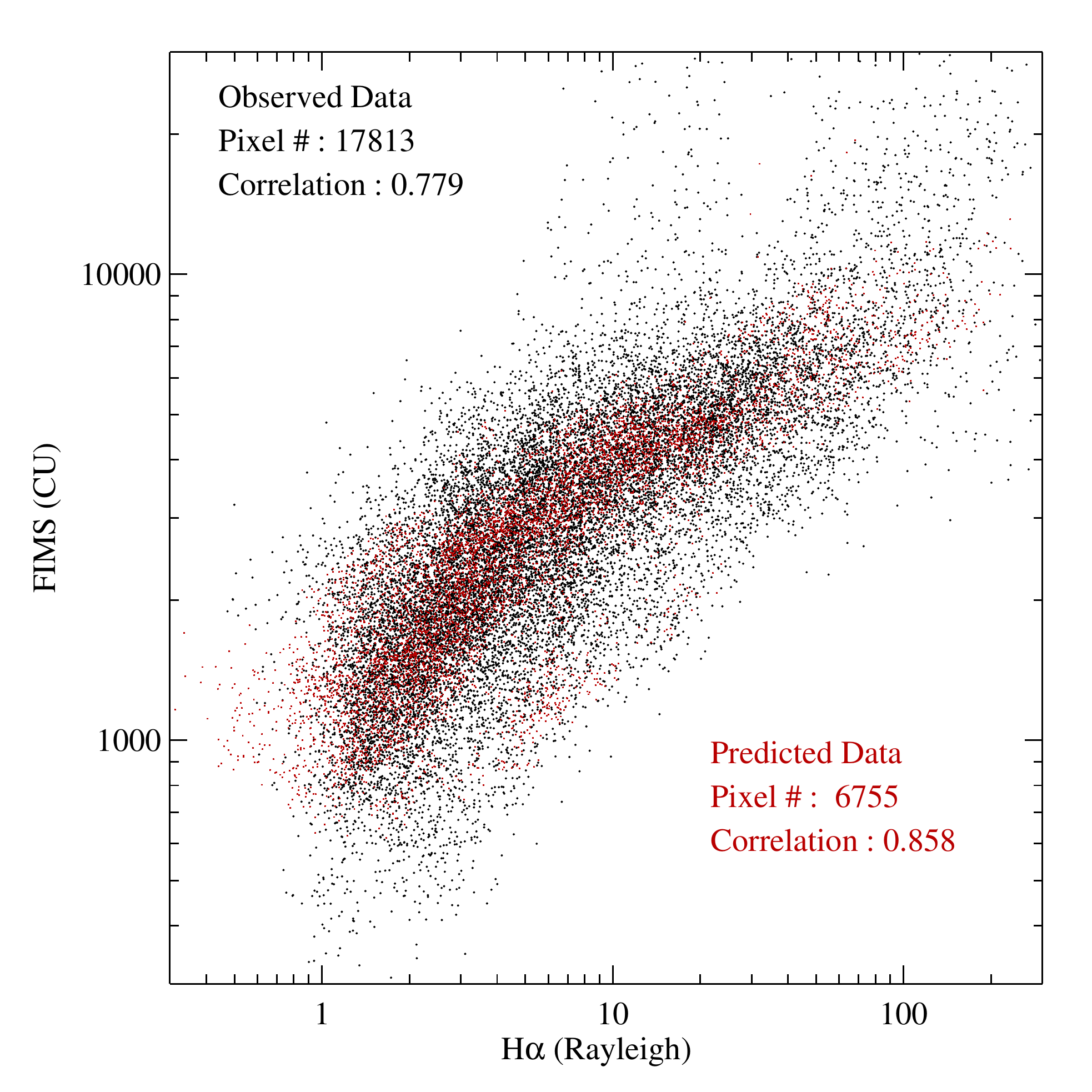} \, \,
		\includegraphics[width=\columnwidth]{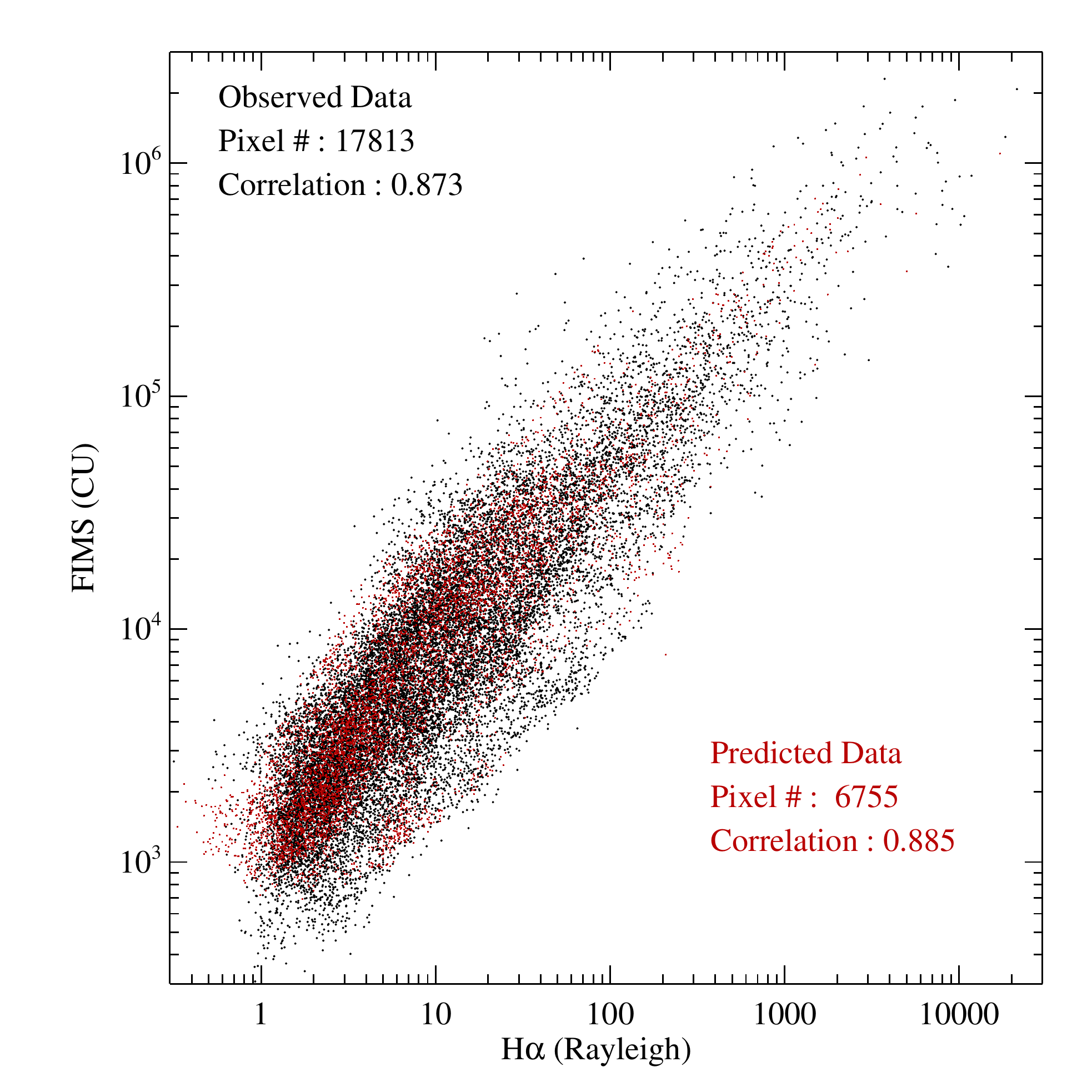}
		\caption{Comparison of the observed (black dots) and predicted (red dots) FIMS intensities with H$\alpha$ intensities for the Galactic plane region of |\textit{b}|<30\degr. Left and right figures show the scatter plots before and after extinction-correction, respectively. }
		\label{fig:f07}
	\end{figure*}
	
	\begin{figure}
		\centering
		\includegraphics[width=\columnwidth]{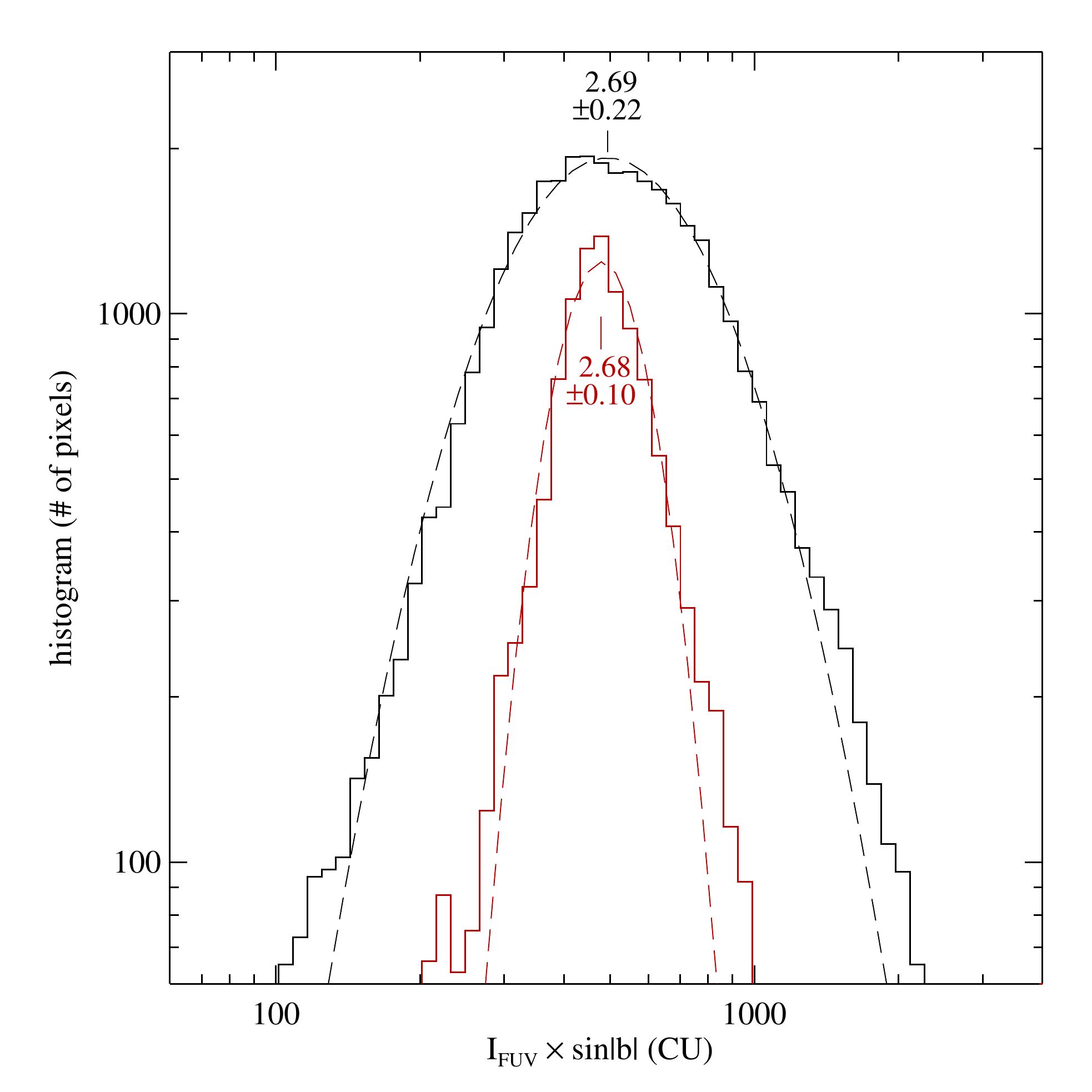}
		\caption{Histograms of I$_{\rm FUV}\times\sin$|\textit{b}| using a logarithmic scale: Black and red histograms correspond to the observed and the predicted FIMS intensities, respectively. Both histograms are fitted with Gaussian curves, with centers and widths indicated.}
		\label{fig:f08}
	\end{figure}

	The comparison shown in Figs. \ref{fig:f04} and \ref{fig:f05} were made only for the regions that were observed by the GALEX. Therefore, we attempted to ensure the validity of the predicted FUV intensities for the bright Galactic plane region, which GALEX did not observe, by employing the following test. We selected a small region of 10$\degr$ < l < 20$\degr$, |b| < 20$\degr$ around the Galactic plane, which FIMS had observed except the lower left corner, and applied the same DNN algorithm to the whole sky with this region intentionally excluded in the learning process. The predicted result for this region was then compared with the actual FIMS observation. Fig \ref{fig:f06} shows the comparison: (a) actual FIMS observation, (b) predicted result, and (c) scatter plot of the predicted intensities compared with the actual observation. As can be seen from the plots, the observed and predicted results are in good agreement, with the correlation coefficient of 0.844 and the mean difference of 17.4$\%$. Though these numbers do not guarantee the reliability of the predicted intensities of the Galactic plane region, we note that the estimated error of FIMS is $\sim$25$\%$ \citep{ede06b}.
	
	Further, we compared the FUV intensities with those of H$\alpha$ for the Galactic low latitude region, because previous studies showed they exhibited a good correlation in the halo region \citep{seo11b}. For the Galactic plane region, where extinction effect is severe, we applied the following simple formula for extinction \citep{nat84,cal94,ben03} with assumption that photon sources and scattering dust grains are uniformly mixed in a plane parallel medium:
	\begin{equation}
		I_{extinction-corrected}=I_{observed}\times\frac{\tau}{1-e^{-\tau}},\label{eq1}
	\end{equation}
	where the optical depth $\tau$ was chosen to be 7.3$\times$E(B-V) for FUV and 2.2$\times$E(B-V) for H$\alpha$ according to the Milky Way dust extinction curve. The correction result was not significantly affected by how the scattering effect was treated (no scattering,  perfect forward scattering, and isotropic scattering). Fig. \ref{fig:f07} shows the observed and predicted FIMS intensities plotted against the H$\alpha$ intensities for the Galactic plane region of |\textit{b}| < 30\degr. The panel on the left is the one before extinction correction. It is apparent that the predicted data points (red dots) closely follow the trends of the observed data points (black dots). Though the data points are widely dispersed, the FIMS FUV intensities correlate strongly with H$\alpha$ intensities for H$\alpha$ intensities less than $\sim$10 R. Above 10 R, FIMS FUV intensities exhibit more extinction than those of visible H$\alpha$. In fact, infrared emission observed at 100 $\mu$m is also bright in this section of the high intensities \citep{seo11b}. The panel on the right is the one with extinction correction, based on the above simple formula. We can see good linear relationship between the FUV and H$\alpha$ intensities, regardless of the level of their intensities. We also note that the observed data points are less dispersed after extinction correction. 
		
	The histograms depicted in Fig. \ref{fig:f08} confirm the validity of the predicted FIMS results. The black and red histograms in Fig. \ref{fig:f08} correspond to the observed and predicted FIMS intensities multiplied by $\sin$|\textit{b}|, respectively. \citet{seo11a} detailed the log-normal property of the diffuse FUV intensity distribution and interpreted it as a natural consequence of the density structure and turbulence property of the ISM. Fig. \ref{fig:f08} shows that the log-normal property is well preserved for the predicted FIMS intensities. While the centers of the two histograms are nearly collocated, the predicted FIMS histogram width is smaller than that of the observed FIMS data, indicating that predicted intensities fluctuate less than the observed ones.

	\begin{figure*}
		\centering
		\includegraphics[clip,scale=0.2]{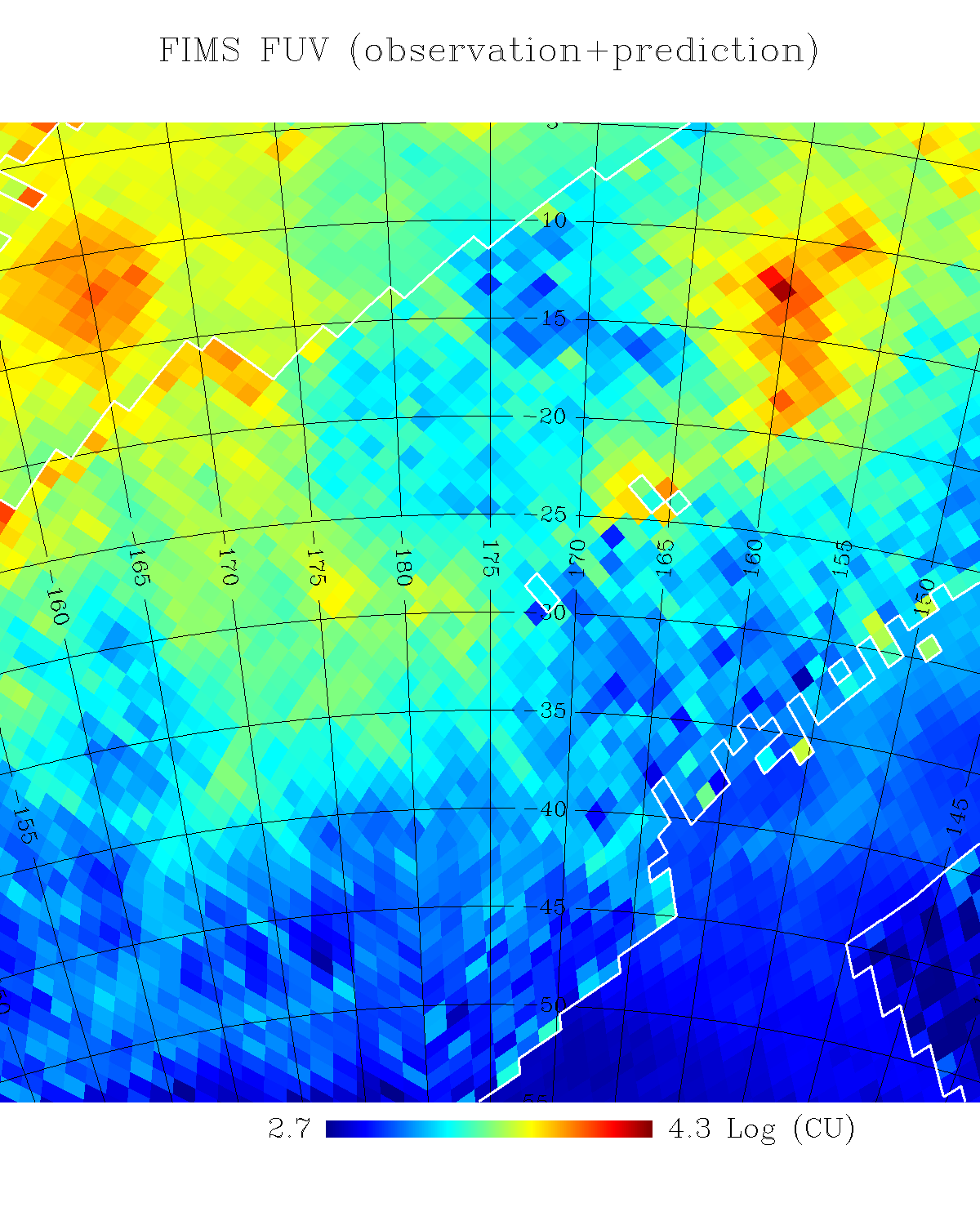} \, \,
		\includegraphics[clip,scale=0.2]{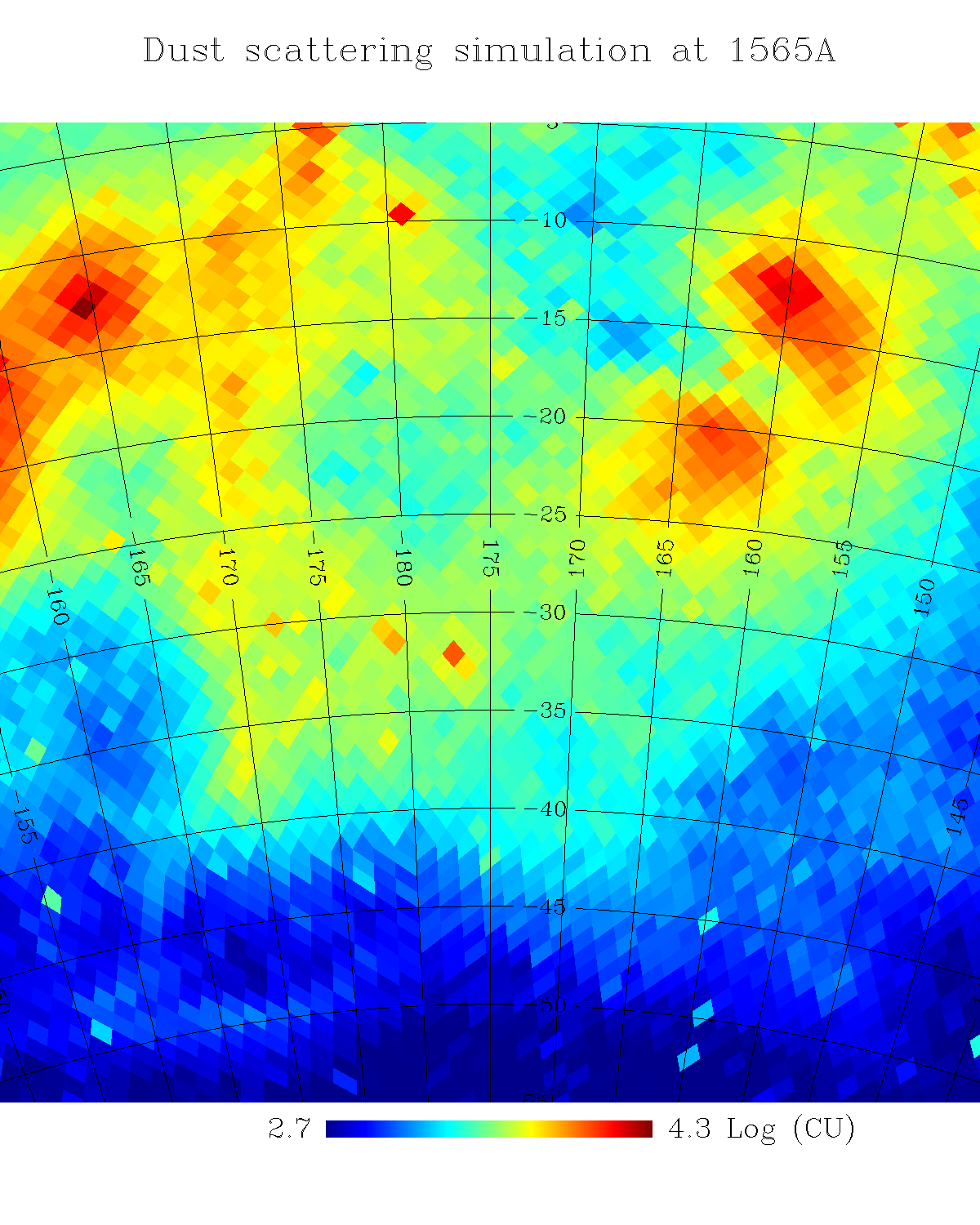}
		\caption{(a) Combined map of the observed FIMS data and the predicted FUV intensities. White contours indicate the boundaries between the observed and predicted data. The large central and small lower-right regions are the observed areas, whereas the upper-left corner and the strip on the lower-right corner are the predicted areas; (b) dust scattering simulation map corresponding to the entire region shown in (a).}
		\label{fig:f09}
	\end{figure*}

	\section{Discussion: Dust scattering simulation}
	\label{sec:discus}
	
	The diffuse continuum background, known as the diffuse Galactic light, comes primarily from starlight scattered by interstellar dust. Here, we use dust scattering simulation results to test the reliability of the DNN-predicted FUV intensities. We selected a relatively large region spanning more than 50$\degr$ in the longitude and latitude directions, centered at (\textit{l}, \textit{b}) = (175\degr, -30\degr), including both observed and predicted regions (Fig. \ref{fig:f09}a). The observed and predicted regions were separated by the white boundaries; the large central region with conspicuous pixel-to-pixel fluctuations is the observed region, and most of the outer regions with smooth intensity are the predicted regions. Dust scattering simulations for the FIMS observations have been reported for two parts of the region shown in the figure, namely the Orion-Eridanus Superbubble \citep{jo12} at the lower left region and the Taurus-Perseus-Auriga Complex \citep{lim13} at the upper right region. The upper left region is Lambda Orionis, analyzed in \citet{lee15} using the data observed by the S2/68 UV telescope \citep{bok73}; neither FIMS nor GALEX observed the region. The  faint lower right region was observed with GALEX but not with FIMS.

	For the simulation’s dust distribution, we combined the three recent 3D dust maps of \citet{lal19}\footnote{Stilism (\href{https://stilism.obspm.fr/}{https://stilism.obspm.fr/} or \href{http://cdsarc.u-strasbg.fr/viz-bin/cat/J/A+A/625/A135}{http://cdsarc.u-strasbg.fr/viz-bin/cat/J/A+A/625/A135})}, \citet{gre19}\footnote{Bayestar19 (\href{http://argonaut.skymaps.info/}{http://argonaut.skymaps.info/})}, and \citet{lei19}\footnote{Galactic dust absorption maps (\href{https://wwwmpa.mpa-garching.mpg.de/~ensslin/research/data/dust.html}{https://wwwmpa.mpa-garching.mpg.de/$\sim$ensslin/research/data/dust.html})} to build a 3D dust map up to a distance of $\sim$2 kpc from the Sun. We took a rectangular box of 800$\times$800$\times$400 cells, with a bin size of 5 pc. Each bin was assigned an extinction density corresponding to the E(B-V) values for 5 parsecs from the E(B-V) value per parsec. Radiation sources were  adopted from the TD-1 \citep{tho78} and Hipparcos \citep{per97} catalogs. The TD-1 catalog provided the observed FUV band fluxes at $\lambda$1565Å, which were then converted to intrinsic luminosities after an extinction correction using updated distances. The Hipparcos catalog provided only the spectral types; therefore, we derived stellar luminosities for the Hipparcos stars using Castelli’s stellar models \citep{cas03} and estimated the FUV fluxes from the stellar luminosities. Distances to the stars were obtained from the GAIA DR2 \citep{gai16,gai18} and Hipparcos catalogs.
	
	Optical properties of dust grains can differ based on line-of-sight and distance. However, previous studies showed that they were similar in the Orion-Eridanus Superbubble and the Taurus-Perseus-Auriga Complex regions with (albedo, g-factor) = ($0.43_{-0.04}^{+0.02}$, $0.45_{-0.2}^{+0.2}$) for the Orion-Eridanus Superbubble region \citep{jo12} and ($0.42_{-0.05}^{+0.05}$, $0.47_{-0.27}^{+0.11}$) for the Taurus-Perseus-Auriga Complex region \citep{lim13}. Therefore, we adopted (albedo, g-factor) = (0.43, 0.45) for our study. A total of 10$^{10}$ stellar photons were used in this simulation.
	
	\begin{figure}
		\centering
		\includegraphics[width=\columnwidth]{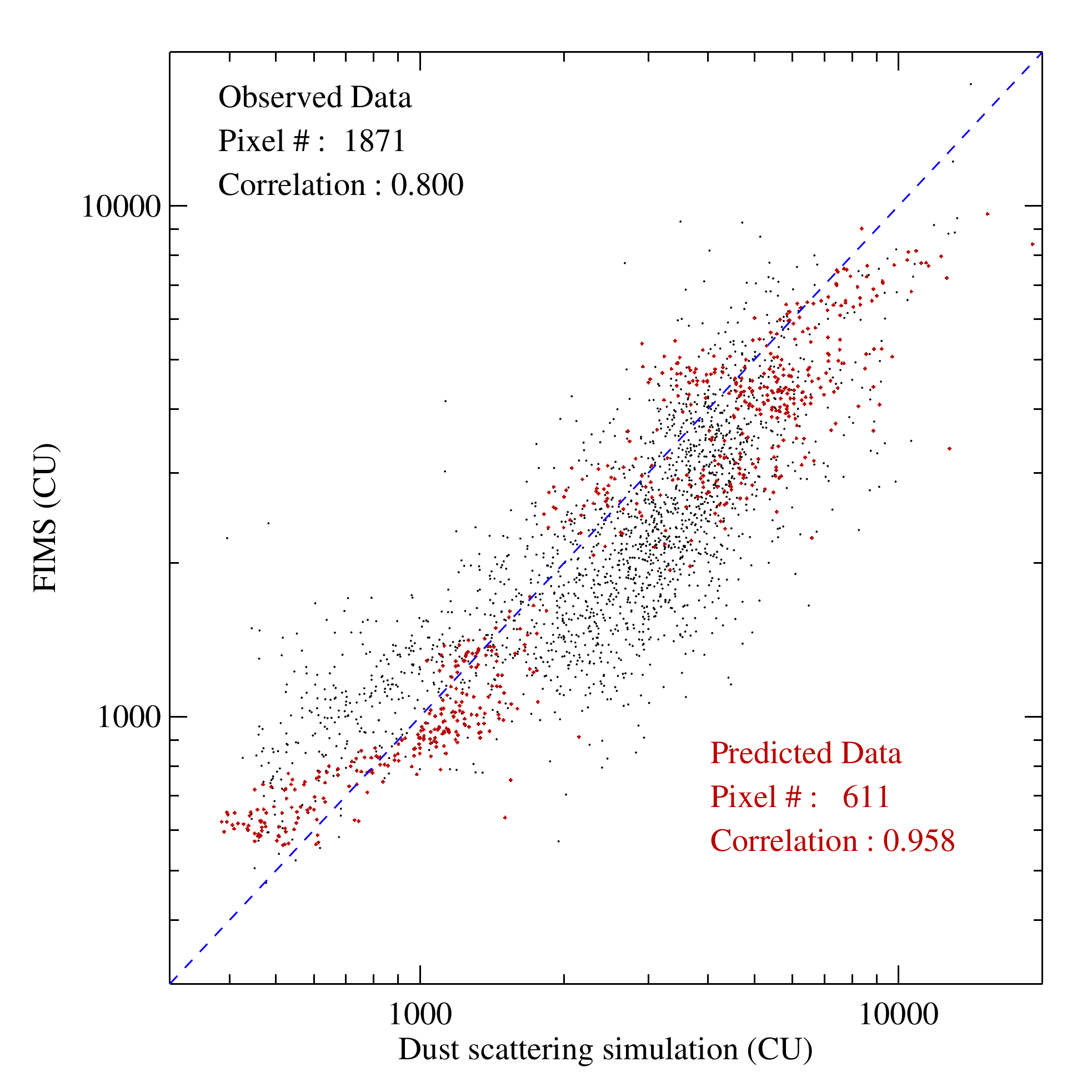}
		\caption{Comparison of the FIMS and dust scattering model intensities. Black and red dots correspond to the observed and predicted data, respectively.}
		\label{fig:f10}
	\end{figure}
	
	\begin{figure}
		\centering
		\includegraphics[width=\columnwidth]{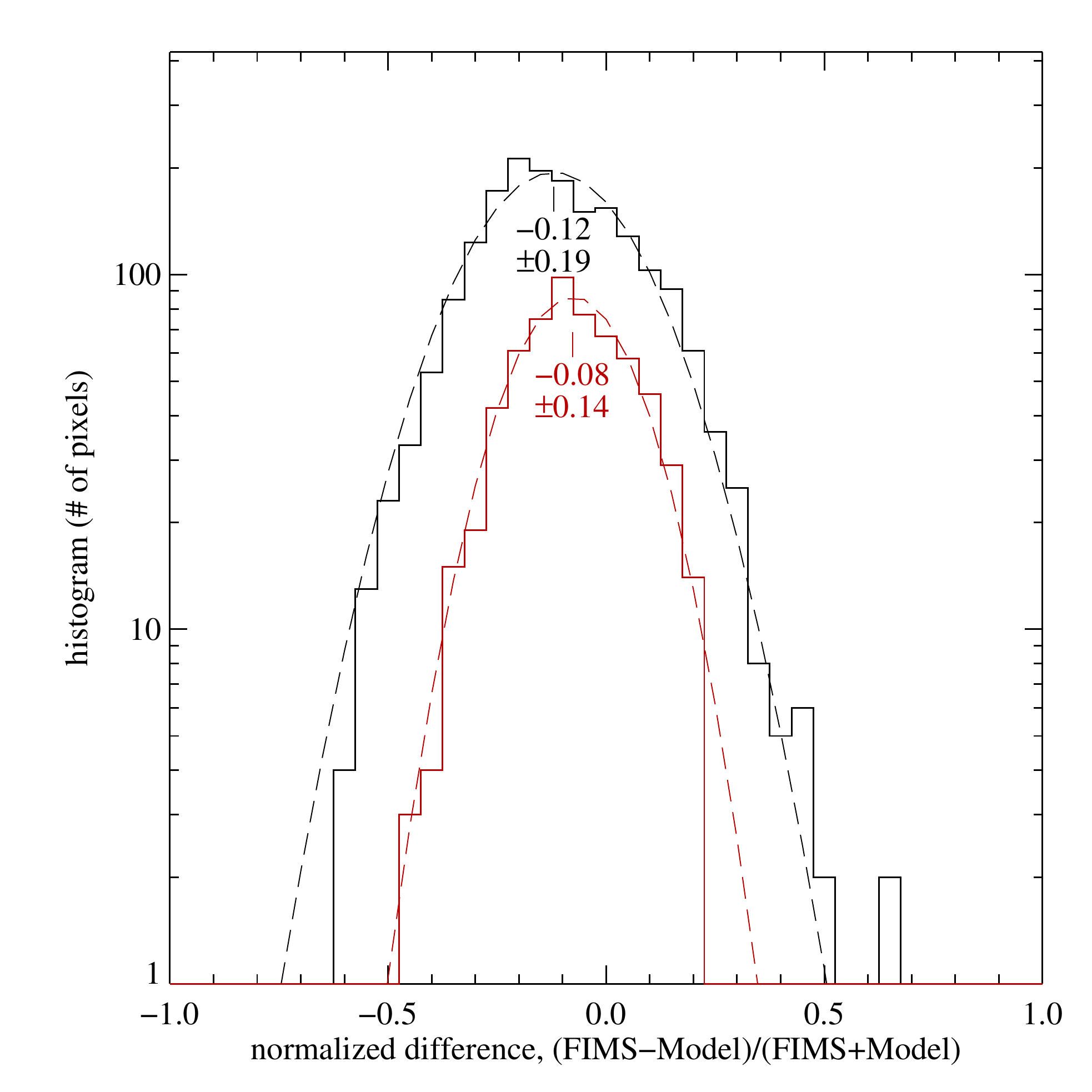}
		\caption{Histogram of differences between the FIMS and the dust scattering model intensities. Black and red histograms correspond to the observed and predicted FIMS data, respectively. Both histograms are fitted with Gaussian curves, with centers and widths indicated.}
		\label{fig:f11}
	\end{figure}

	Fig. \ref{fig:f09}(b) shows simulation results (henceforth the model map) using the same pixel size and color scheme as in Fig. \ref{fig:f09}(a). The model map exhibits good agreement with the combined map, although the simulation map is smoother in part due to lower resolution of the 3D dust map. Furthermore, the overall intensity variation is larger in the model map than in the combined map, possibly because the model dust map has extinction densities lower than their actual values and no high enough spatial resolution to represent the complex density variation of dust clouds. In fact, the extinction density up to 100 pc from the Sun is much smaller in the 3D dust map of \citet{lei19} than in other maps, which lowers the dust-scattered intensity. It is also known that a clumpy medium has a lower effective optical-depth compared to a homogeneous medium with the same dust mass and yields a lower scattered light in high-intensity regions and a higher scattered light in low-intensity regions \citep[e.g.,][]{seo13,seo16}. The excess of the model intensity in high-intensity regions is likely to be attributable to both effects. The expectation in a clumpy medium explains the general trend shown in Fig. \ref{fig:f09} very well. We also note that \citet{lee15} obtained slightly different albedo and g-factor of $(0.31^{+0.03}_{-0.03}, 0.65^{+0.06}_{-0.06})$ using a shell + ring model, which cannot be represented with a 5-pc spatial resolution, for the Lambda Orionis nebula. This difference stresses the importance of accurate knowledge of the 3D dust density.
	
	The simulation model’s validity is further analyzed in Fig. \ref{fig:f10}, in which the FIMS observation and prediction pixels are compared with those of the simulation. In addition, Fig. \ref{fig:f11} shows two histograms of the differences between the observed/predicted intensities and the dust scattering simulations ((I$_{\rm FIMS}$-I$_{\rm Model}$)/(I$_{\rm FIMS}$+I$_{\rm Model}$)). It is impressive that the 0.96 correlation coefficient for the predicted data is larger than the coefficient 0.80 for the observed data. The higher correlation coefficient for the predicted map may be due to its smaller pixel-to-pixel statistical fluctuations. Both the observed and predicted maps exhibit $\sim$300 CU higher intensities than the model simulation map in the low intensity region of < $\sim$1000 CU. This $\sim$300 CU excess may be due to extragalactic background emission \citep{gar00,bro00}, and/or an inaccurate representation of the detailed density structure in the 3D dust map. Fig. \ref{fig:f11} also indicates high confidence in the predicted FIMS intensities. The statistical deviations of the simulation results from the model values have similar Gaussian distributions for both the observed and the predicted regions. Consequently, the dust scattering simulation well reproduces the overall feature of the observed and DNN-predicted FIMS data. This result suggests that the DNN algorithm provides a reasonably good prediction for unsurveyed FUV sky map.

	\section{Summary}
	\label{sec:summa}
	
	In this study, a deep learning algorithm was applied to fill unsurveyed regions of the FUV all sky survey and a combined map was constructed based on the observed and predicted intensities for the whole sky map. The all sky maps of H$\alpha$, E(B-V), N(HI), and two X-ray bands, together with Galactic longitudes and latitudes, were employed as input parameters while the incomplete FUV intensity map made by FIMS was a target parameter. Even with a simple four-layer neural network architecture that consisted of three convolution layers and a final dense layer, results were remarkably consistent with other observations. To test the validity and usefulness of the predicted intensities further, a small region of the combined map, including both the observed and the predicted regions, was compared to the dust scattering simulation model that utilized recent 3-D Galactic dust maps and star catalogs (e.g., TD-1 and Hipparcos). Simulation results exhibited good agreement with the combined FUV map in the observed and predicted regions.
	
	Whereas the predicted FUV intensities need to be confirmed by direct observations, the present combined map could be a starting point for studies that involve the FUV continuum emissions. This includes the extinction/scattering studies for interstellar clouds, as illustrated in Section \ref{sec:discus}. The FUV all sky map also provides important clues to the radiation transfer mechanisms on a global scale, especially when used together with the global maps in other wavelengths. Further, the optically thick and complex Galactic low latitude regions are interesting and challenging in view of deep learning. Finally, we note here that the final combined map, based on the HEALPix scheme with an angular resolution of N$_{\rm side}$ = 64, can be downloaded from the website\footnote{\href{https://drive.google.com/file/d/1841qiiVj4E49kZFkNU3aAhvbbZ3qxMBM/view?usp=sharing}{https://drive.google.com/file/d/1841qiiVj4E49kZFkNU3aAhvbbZ3qxMBM/view?usp=sharing}}. 
	 
	\section*{Data Availability}
	The data underlying this article are available in the GitHub link of “https://github.com/yjchoi83/FIMS\_FUV”.

	\section*{Acknowledgements}
	
	This research was supported by the Korea Astronomy and Space Science Institute under the R\&D program supervised by the Ministry of Science, ICT, and Future Planning of Korea. Support was also provided by the Basic Science Research Program through the National Research Foundation of Korea (NRF), funded by the Ministry of Education (2019R1F1A1061102).

	
	
	


	\appendix
	
	\section{Test of the impact of an unreasonable input dataset}
	
\begin{figure}
		\centering
		\includegraphics[clip,scale=0.33,angle=90]{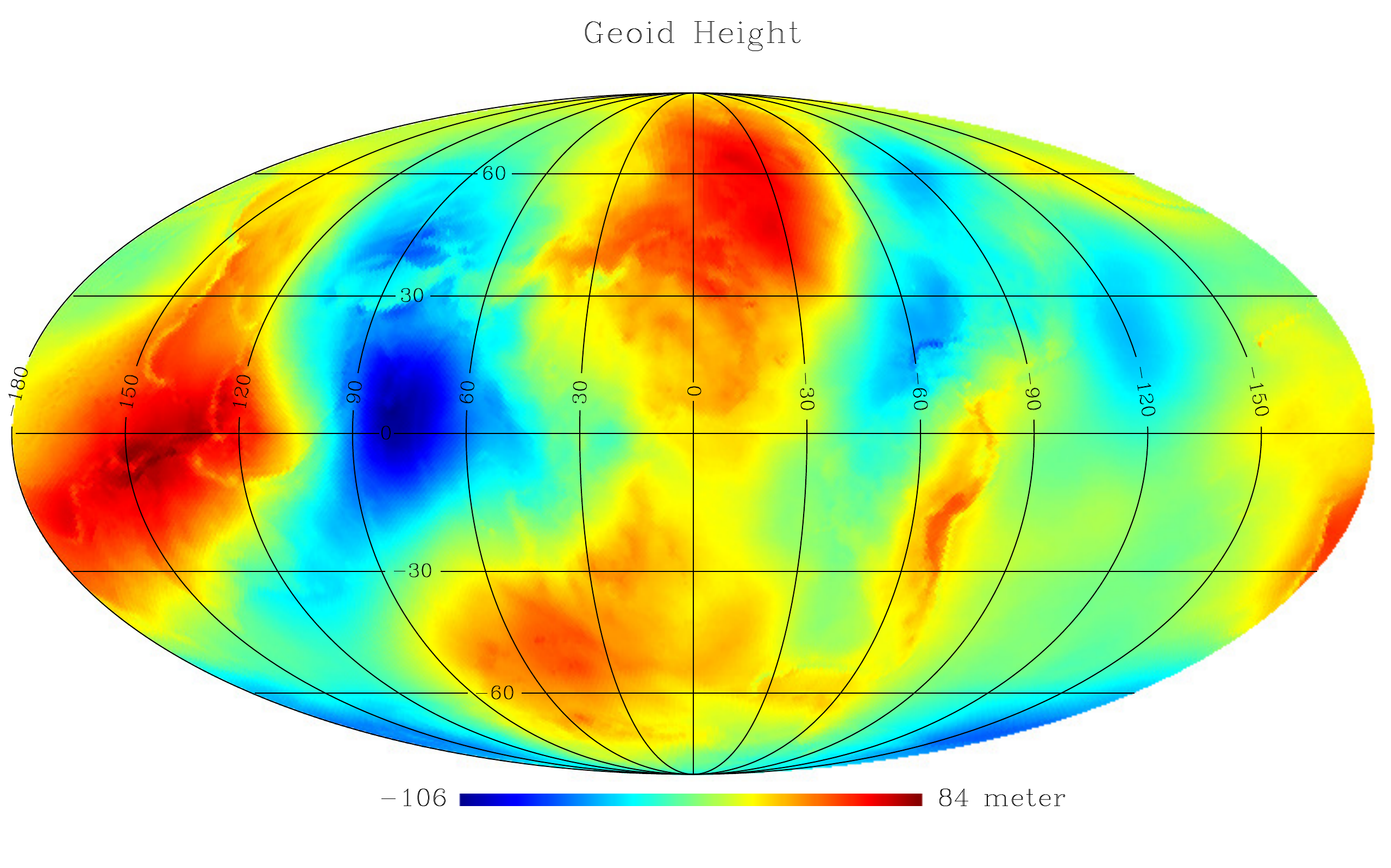}\\
		\includegraphics[clip,scale=0.33,angle=90]{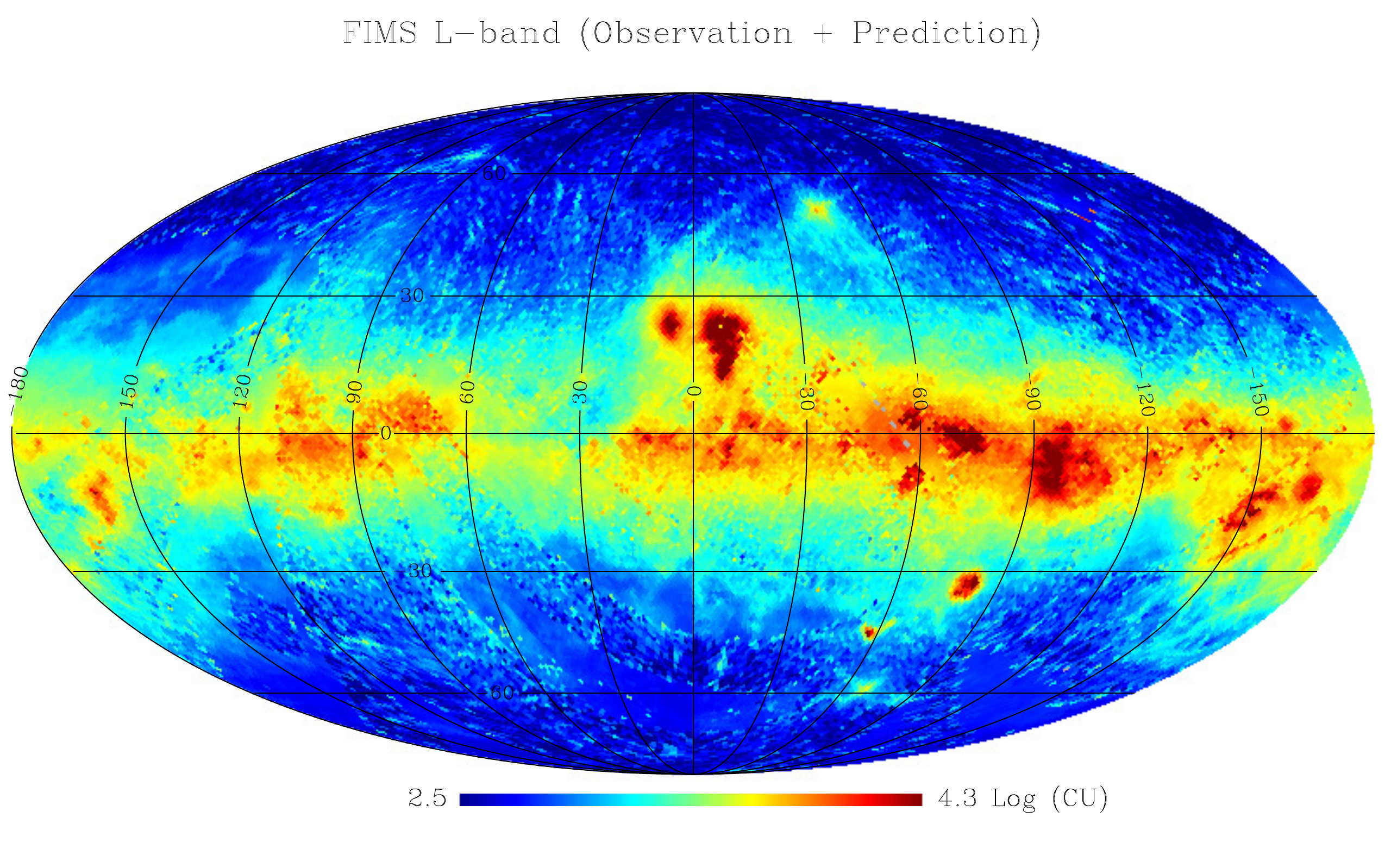}\\
		\includegraphics[width=\columnwidth]{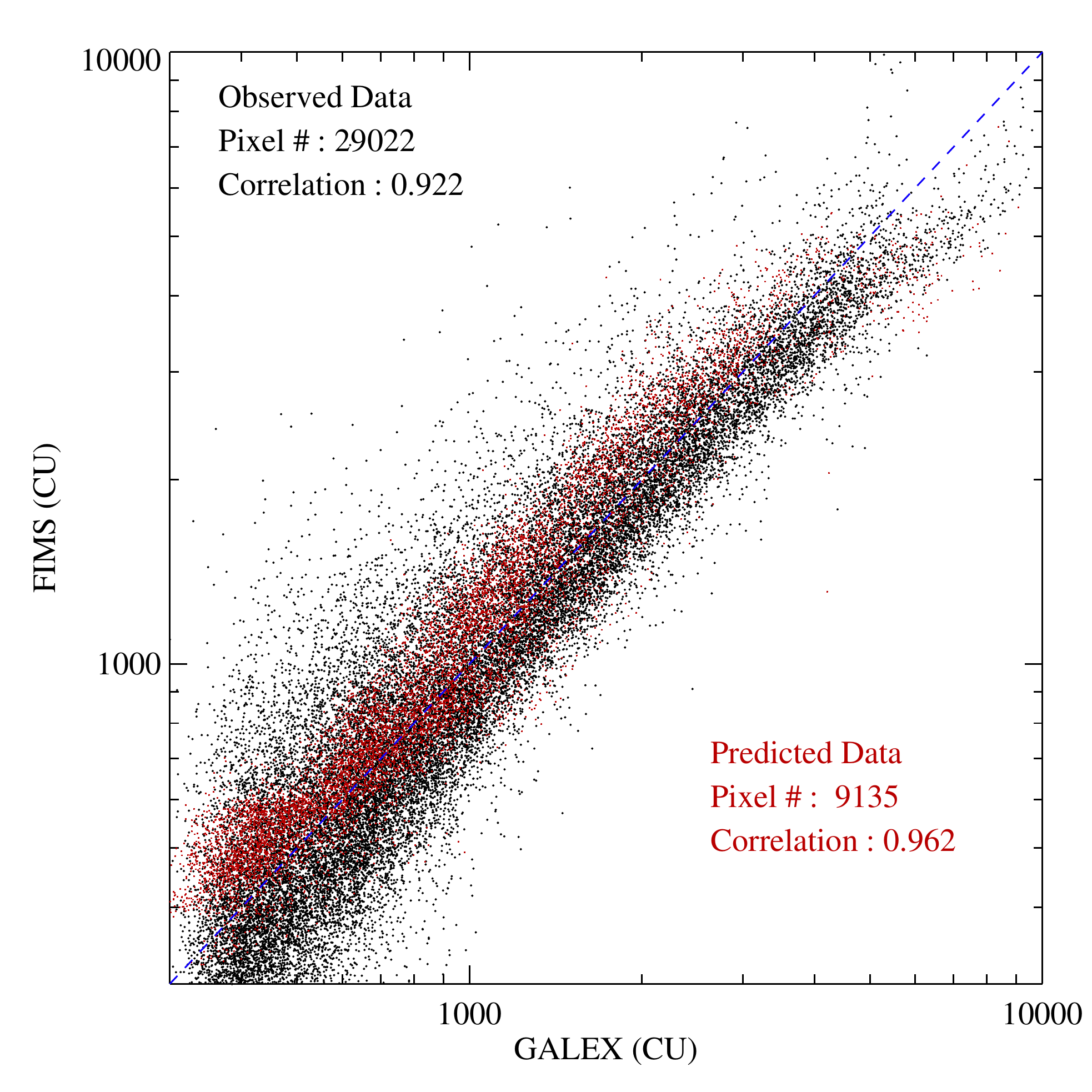}\\
		\caption{(a) Geoid height map (NGA/NASA EGM96, N=M=360 Earth Gravitational Model), adopted as the eighth input dataset to test the impact of an irrelevant input, (b) combined map of the resulting FUV predictions and the observations, corresponding to Fig. \ref{fig:f03}, and (c) comparison of the observed and predicted FIMS intensities with GALEX intensities, corresponding to Fig. \ref{fig:f04}.}
		\label{fig:fA01}
	\end{figure}

	We argued in the main text that the final prediction map in Fig. \ref{fig:f03} is reasonable through various tests. However, it should be noted that successful prediction depends on a proper choice of input datasets; however, they can also be arbitrary. One trivial question in this line is, how would a totally irrelevant input dataset affect the result? To answer this question, we tested the impact of an unreasonable input dataset with a geoid height map\footnote{\href{https://earth-info.nga.mil/GandG/wgs84/gravitymod/egm96/egm96.html}{https://earth-info.nga.mil/GandG/wgs84/gravitymod/egm96/egm96.html}}, which has nothing to do with the present FUV study. We took the geoid height map, shown in Fig. \ref{fig:fA01}a, as the eighth input dataset and ran the same DNN code, along with the seven original datasets. The result is shown in Fig. \ref{fig:fA01}b, which is hardly different from the original result in Fig. \ref{fig:f03}. We also compared the predicted FUV intensities with the corresponding GALEX observations in Fig. \ref{fig:fA01}c and obtained a result very similar to that shown in Fig. \ref{fig:f04}. We believe that this test confirms that an irrelevant dataset would not contribute to the final result, implying that the input datasets adopted in this study contribute to the final predicted FUV map with appropriate weights.
	
	\bsp	
	\label{lastpage}
\end{document}